\documentclass{ifacconf}

\usepackage{graphicx}      % include this line if your document contains figures
\usepackage{natbib}        % required for bibliography
\usepackage{amssymb}
\usepackage{amsmath}
\usepackage{placeins}
\usepackage{xcolor}

\begin{document}
	\begin{frontmatter}
		
		\title{A two dimensional fluid model for TCP/AQM analysis\thanksref{footnoteinfo}} 
		
		\thanks[footnoteinfo]{A.J. Rojas is grateful for the support of ANID, through Basal Project AC3E grant FB0008. \\ This work has been submitted to IFAC for possible publication.}
		
		\author[First]{BELAMFEDEL ALAOUI, Sadek} 
		\author[Second]{ROJAS, Alejandro J.} 
		\author[Third]{HMAMED, Abdelaziz}
		\author[First]{TISSIR El Houssaine}
		
		\address[First]{Laboratory of Informatic, Signals, Automatic and Cognitivism, Faculty of sciences Dhar el mehraz University Sidi Mohamed Ben Abdellah (e-mails: sadek.belamfedelalaoui, elhoussaine.tissir (at)usmba.ac.ma).}
		\address[Second]{Departamento de Ingenieria Electrica, Universidad de Concepcion, Concepcion, Chile}
		%		\address[Fourth]{Electrical Engineering Department, 
		%			Seoul National University, Seoul, Korea, (e-mail: author@snu.ac.kr)}
		\address[Third]{Campus de l'université privée de Fès (hmamed(at)upf.ac.ma) }
		\begin{abstract}                % Abstract of not more than 250 words.
			This work proposes a new mathematical model for the TCP/AQM system that aims to improve the accuracy of existing fluid models, especially with respect to the sequential events that occur in the network. The analysis is based on the consideration of two time bases, one at the queue's router level and the other at the congestion window level, which leads to the derivation of a new nonlinear two-dimensional fluid model for Internet congestion control. To avoid the difficult task of assessing stability of a 2D nonlinear dynamic model, we perform a local stability analysis of a 2D linear TCP AQM model. By constructing a new two dimensional second order Bessel Legendre Lyapunov functional, new matrix inequalities are derived to evaluate the stability of the 0-input system and to synthesize a feedback controller. Finally, two Internet traffic scenarios, with state space matrices for replicability,  are presented, demonstrating the validity of the theoretical results.
		\end{abstract}
		
		
	\end{frontmatter}
	%===============================================================================
	
	\section{Introduction:} %murak
\vspace{-5pt}	Active Queue Management (AQM) addresses the bufferbloat issue \cite{staff2012bufferbloat}. This algorithm actively interacts with congestion control algorithms, in particular the Transmission Control Protocol (TCP) but also for any other transport protocol, to send higher amounts of data through the network. The buffer space available in routers and switches should meet the short-term buffering requirements. AQM schemes aim to reduce buffer occupancy and, as a result, end-to-end delay, see the Controlled Delay (CoDel) scheme \cite{nichols2012controlling} and the Proportional Integral controller Enhanced (PIE) scheme \cite{pan2013pie}.  
	
	There are several mathematical models for TCP/AQM fluid dynamics: \cite{mathis1997macroscopic,kelly1998rate,low2003duality,misra2000fluid,xu2015new}. The model proposed in \cite{mathis1997macroscopic} introduces significant inaccuracies due to many simplifications. The models proposed in \cite{kelly1998rate} and \cite{low2003duality} are not scalable. The MGT model proposed in \cite{misra2000fluid} and its simplified version \cite{hollot2001control}, are the most used models for AQM synthesis. It was shown in \cite{xu2015new}, that the MGT model does not describe the behavior of the TCP/AQM loop in various network scenarios. The article \cite{xu2015new} proposed a model that takes into account different network scenarios. Several extensions and improvements of the model of \cite{xu2015new} have been developed in the literature, among which are, \cite{BELAMFEDELALAOUI201858} for multibottleneck topologies with successive delays and \cite{alaoui2019modelling,belamfedel2021new} that model the effect of a denial of service attack over a network.
	
	\textbf{Motivation:} This work is mainly motivated by the challenges arising in the design of efficient active queue management (AQM) schemes. There have been many results in the past few years on the synthesis of AQMs rendering the TCP/AQM closed-loop system stable and performing, see for instance \cite{sadek2019small,sadek2020congestion,kar2022paqman}, and references therein. All previous approaches explored in the literature have considered the TCP/AQM system from a one-dimensional time base, \cite{misra2000fluid,xu2015new,BELAMFEDELALAOUI201858,vardoyan2018towards,domanski2020diffusion,jing2022multiple}, which is limited since it does not respect the temporal sequential events that occur in the network. 	
	
\begin{figure}
		\centering
		\includegraphics[width=1\linewidth]{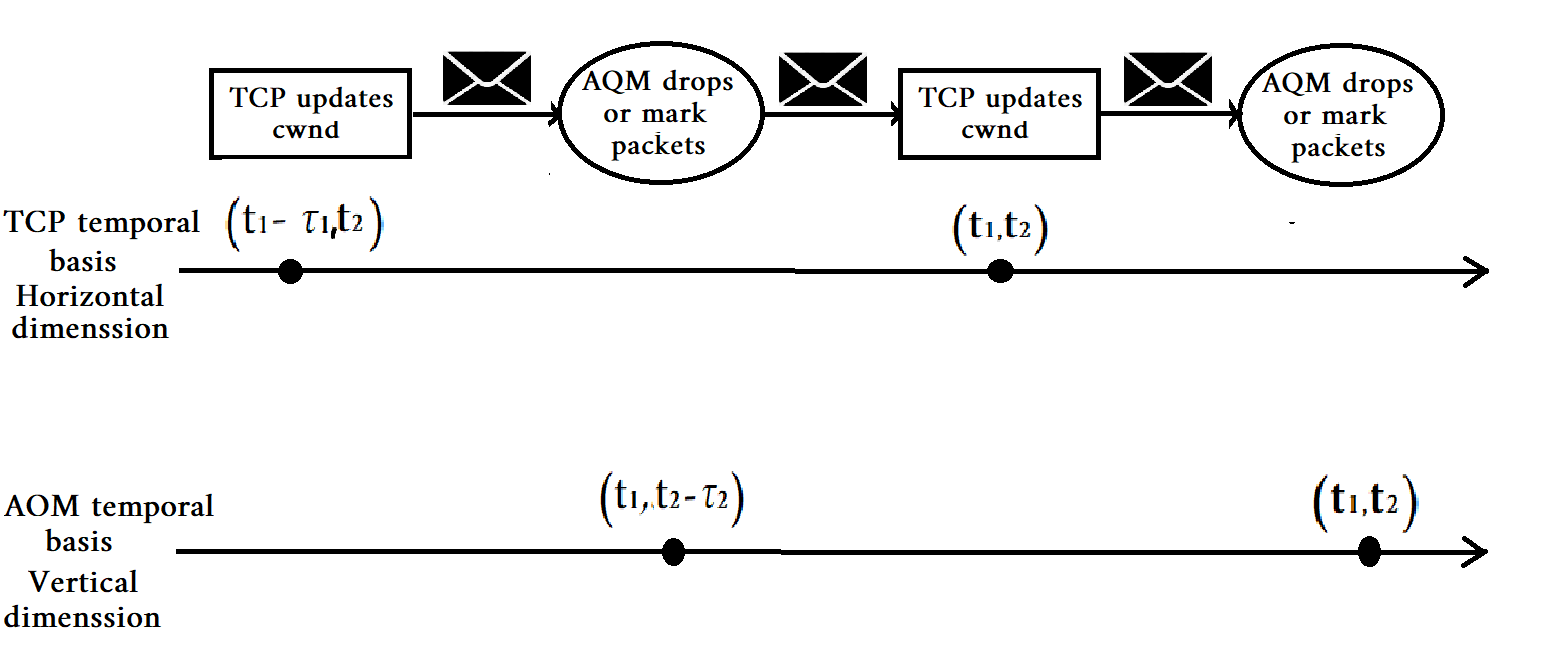}
		\caption{Sequential events of data transfer in Internet}
		\label{fig:2dtcpaqm}
	\end{figure}

	To overcome this limitation, we introduce a novel accurate two-dimensional (2D) fluid model for TCP/AQM analysis. The model is designed from two temporal basis points of view, one at the router level and the other at the server level. From an accurate temporal sequence  of data transfer (see Fig. \ref{fig:2dtcpaqm}), we derive two dimensional differential equations.  The resultant framework offers opportunities to analyze the stability of a diverse set of controllers that could be linear or nonlinear in one  or two dimensional spaces. First, we show that the new model is more general and can be reduced to the one dimensional model in \cite{xu2015new}.  We then analyze the proposed 2D fluid model and find the system's unique equilibrium point, for which we  deduce a linear approximation of the model using the first order Taylor expansion around it. Since all the resultant partial derivatives that comprise the linear process are continuous Lipschitz functions, the linear process's stability analysis implies the nonlinear process's local stability \cite{gu2003stability}. At this stage, we construct for the first time a two dimensional second order Bessel Legendre Lyapunov functional. This latter permits to derive first a less conservative LMI condition for stability assessment of 2D time delay systems, and second a result for feedback gain synthesis. Finally, some simulations using MATLAB shows that the feedback controller achieves a stabilisation to the solution of the 2D system.
	
	\textbf{Notation :} Over this paper, $W(t)\in \left[ 0, \bar{W}\right] $ is the average congestion window size in packets, $\tau(t) = q(t)/C(t) + Tp$ is the round trip time $ RTT $ and $p(t)\in \left[ 0, 1\right] $ is the probability of dropping/marking packets. $q\in \left[ 0, Q_{max}\right]$, $C$ and $T_{p}$ denote the queue length (packet), link capacity (packet/s) and propagation delay $(s)$, respectively. $ \lambda $ is the window distribution parameter. It is a function of $ N $ and the congestion window $ (cwnd) $ and it satisfies $ 1\leq \lambda \leq N $. For large $ N $, the extreme cases of $ \lambda = 1 $ and $ \lambda = N $ are very rare, \cite{xu2015new}. $ ssthresh $ denotes the Slow Start Threshold. ECN for Explicit Congestion Notification. $ Z \in \mathbb{S}^{n} $ means that the matrix $ Z $ is a symmetric matrix with size $ n\times n $.  $ \mathbb{R}^{n} $ denotes the n-dimensional Euclidean space. The notations $ \mathbb{R}^{n\times m} $ and $ \textbf{S}^{n} $ are the set of $ n \times m $ real matrices and of $ n \times n $ real symmetric matrices, respectively. The notation $ P\in \mathbb{S}_{+}^{n} $, means that $ P \in \mathbb{S}^{n} $ and $ P > 0 $, which means that P is symmetric positive definite. The symmetric elements of a symmetric matrix will be denoted by $ \ast $. For any matrices $ X,Y $ of appropriate dimension, the matrix $ diag(X,Y) $ stands for $ \left[\begin{array}{cc}
		X & 0 \\
		0 & Y
	\end{array}\right] $.
	The matrices $ I_{n} $ and $ 0_{n,m} $ represent the identity and null matrices of appropriate dimension and, when no confusion the subscript will be omitted. Moreover, for any square matrix $ Z $, we define $ He(Z) = Z+Z^{T} $. $ x_{t}(r) $ and $ \dot{x}_{t} $ denote $ x(t+r) $ and $ \dot{x}(t+r) $, respectively.
	
	\section{two dimensional fluid model for TCP/AQM analysis}
	%	This section introduces the TCP/AQM system and explains the limitations of the one dimensional models existing in the literature, e.g, \cite{xu2015new}. Upon this limitation, a novel two dimensional fluid model for TCP/AQM analysis is developed.
	
	In Internet, routers must be able to buffer packets. Unfortunately, there is no guarantee that buffers are only used for buffer transients. In particular, TCP congestion control increases the amount of data on the network each round trip in order to maintain high link utilization, but once the path from source to destination is full, excess packets are queued at the upstream end of the slowest link in the path \cite{}. Since today's node interfaces can buffer hundreds or thousands of packets, long-lived connections can create many seconds of unnecessary delay. AQM within routers in Internet aims at reducing the queueing delay while maximizing the transmission rate of TCP sources. AQM describes methods for sending proactive feedback to TCP flow sources in order to regulate their throughput by dropping or marking packets.

	\subsection{Limitation of the one dimensional model}
	The main limitation behind the one dimensional TCP AQM model is that when TCP approximates the Quality of Service (QoS) of Internet it uses its own temporal basis. On the other hand, when the AQM acts on the queue it acts on another temporal basis which is different from the temporal basis of TCP. To precisely explain this understanding, let us consider the one-dimensional dynamic equation of TCP in \cite{xu2015new}, scenario A,
	\begin{align} \label{XuA}
		\resizebox{0.98\hsize}{!}{$ \dot{W} (t)=\frac{W(t-\tau)}{\tau(t-\tau)} (1-p(t-\tau))-\lambda \frac{W(t)W(t-\tau)}{2N\tau(t-\tau)} p(t-\tau). $}
	\end{align}
	Details on the name of each variable are given in the notation paragraph. Considering a single time base, the quantity $ W(t-\tau) $, where $ t $ represents the time base at the TCP source, denotes the value of the congestion window, at the time the packets are sent $ t-\tau $, before the round trip time. The signal $ p(t-\tau) $ represents the associated probability of regulating the congestion window at time $ t-\tau $. The problem here is that the packets have just been sent and so this drop or mark probability cannot be applied to packets sent at time $ t-\tau $. We emphasize here that this limitation refers not only to the \cite{xu2015new} model but also to other models in the literature, \cite{misra2000fluid,BELAMFEDELALAOUI201858,vardoyan2018towards,domanski2020diffusion,jing2022multiple}.
	
	\subsection{Congestion window in two dimensional space}
	
	Our approach to the TCP/AQM system overcome this problem in the modelling and consider two temporal basis. One at the TCP edge denoted $ t_{1} $ (called horizontal dimension), which is used to reflect the dynamic motion of TCP seen from the TCP temporal basis. The second at the router denoted $ t_{2} $ (called vertical dimension), which is used to reflect the dynamic motion of TCP seen from the router temporal basis. Since we have two time references $ t_{1} $ and $ t_{2} $, the dynamic motion of the congestion window function is modelled on both temporal dimensions. Here, we define the following signals,
	\begin{itemize}
		\item[$\bullet$] $ W^{h}(t_{1},t_{2}) $: The effective released average congestion window (the real congestion window known by the TCP source) at the horizontal time $ t_{1} $ that will be subjected to drop/mark at the vertical time $ t_{2} $.
		\item[$\bullet$] $ W^{v}(t_{1},t_{2}) $: The average congestion window measured on the queue side by the rate mismatch at time $ t_{2} $ corresponding to the effective average congestion window $ W^{h}(t_{1},t_{2}) $ released at time $ t_{1} $.
		\item[$\bullet$] $ p^{h}(t_{1},t_{2}) $: the probability that, on the part of the TCP, has abandoned or marked its effective average congestion window $ W^{h}(t_{1},t_{2}) $. 
		\item[$\bullet$] $ p^{v}(t_{1},t_{2}) $: The effective drop/mark probability at time $ t_{2} $ corresponding to the flow released at the horizontal time $ t_{1} $. 
		\item[$\bullet$] $ q^{h}(t_{1},t_{2}) $: The queue length obtained from the arrival flow $ W^{h}(t_{1},t_{2}) $.
		\item[$\bullet$] $ q^{v}(t_{1},t_{2}) $: The effective queue length obtained from the arrival flow $ W^{v}(t_{1},t_{2}) $.
	\end{itemize}
	Fig. \ref{fig:2dtcpaqm} illustrates the TCP/AQM system cycle and indicates the temporal evolution with respect to the two temporal dimensions. 
	
	\underline{Motion of the cwnd in the horizontal dimension:} It is first worth noticing that the model in \cite{xu2015new} takes into account the three 3 modes of TCP,
	\begin{description}
		\item[\textit{\underline{Slow Start:}}] The Slow start gradually increases the amount of data transmitted until it finds the network's maximum carrying capacity. Based on this value, it determines a certain threshold called Slow Start Threshold (ssthresh). After this mode, TCP switches to  Congestion Avoidance mode.
		\item[\textit{\underline{Congestion avoidance:}}] This mode differs from a TCP variant to another. For the new reno variant considered in this paper, TCP follows the additive increase multiplicative decrease algorithm.
		\item[\textit{\underline{Fast recovery:}}] When an acknowledgment (ACK) is duplicated, rather than reverting to Slow Start mode (and after going through fast retransmit mode), TCP resend the lost segment and wait for an ACK for the entire previously transmitted window before returning to Congestion Avoidance mode.
	\end{description}
	
	To derive the dynamic motion of TCP on horizontal dimension, we must place ourselves on the time axis of TCP at the time TCP updates its cwnd. On the horizontal time $ t_{1} $ the cwnd $ W^{h}(t_{1}, t_{2}) $ is updated based on the cwnd at $ (t_{1}-\tau_{1}(t_{1}), t_{2}) $ and the drop mark action performed at $ (t_{1}, t_{2}-\tau_{2}) $. 
	
	\begin{description}
		\item[Scenario A:] All the TCP sessions $ W^{h}(t_{1},t_{2}) < ssthresh $, so they are either in slow start, or in fast recovery.	
	\end{description}
	
	\begin{align}
		\label{Eq2}\frac{\partial W^{h}\left(t_{1}, t_{2}\right)}{\partial t_{1}}=&\frac{W^{h}(t_{1}-\tau_{1},t_{2})}{\tau_{1}(t_{1}-\tau_{1})} (1-p^{v}(t_{1},t_{2}-\tau_{2}))  \nonumber\\
		&-\lambda \frac{W^{h}(t_{1},t_{2})W^{h}(t_{1}-\tau_{1},t_{2})}{2N\tau_{1}(t_{1}-\tau_{1})} p^{v}(t_{1},t_{2}-\tau_{2}), 
	\end{align}
	\begin{description}
		\item[Scenario B:] All the TCP sessions $ W^{h}(t_{1},t_{2}) \geq ssthresh $, so they are either in congestion avoidance, or in fast recovery.	
	\end{description}
	\begin{align} \label{Eq3}
		\frac{\partial W^{h}\left(t_{1}, t_{2}\right)}{\partial t_{1}}=&\frac{NW^{h}(t_{1}-\tau_{1},t_{2})}{\tau_{1}(t_{1}-\tau_{1})W^{h}(t_{1},t_{2})} (1-p^{v}(t_{1},t_{2}-\tau_{2})) \nonumber\\
		&-\lambda \frac{W^{h}(t_{1},t_{2})W^{h}(t_{1}-\tau_{1},t_{2})}{2N\tau_{1}(t_{1}-\tau_{1})} p^{v}(t_{1},t_{2}-\tau_{2}),
	\end{align}
	
	\underline{Motion of the cwnd in the vertical dimension:} To derive the dynamic motion of TCP on vertical dimension, we must place ourselves on the AQM temporal basis at the time the AQM performs its action.
	
	\begin{description}
		\item[Scenario A:] All the TCP sessions $ W^{h}(t_{1},t_{2}) < ssthresh $, so they are either in slow start, or in fast recovery.	
	\end{description}
	\begin{align} 
		\label{Eq4} \frac{\partial W^{v}\left(t_{1}, t_{2}\right)}{\partial t_{2}}& = \frac{W^{h}(t_{1},t_{2})}{\tau_{2}} (1- p^{v}(t_{1},t_{2}) \nonumber\\
		&-\lambda \frac{(W^{h}(t_{1},t_{2}))^{2}}{2N\tau_{2}} p^{v}(t_{1},t_{2}),
	\end{align}
	
	\begin{description}
		\item[Scenario B:] All the TCP sessions $ W^{h}(t_{1},t_{2}) \geq ssthresh $, so they are either in congestion avoidance, or in fast recovery.	
	\end{description}
	\begin{align} \label{Eq5}
		\frac{\partial W^{v}\left(t_{1}, t_{2}\right)}{\partial t_{2}}&=\frac{NW^{h}(t_{1},t_{2})}{\tau_{2}} (1-p^{v}(t_{1},t_{2})) \nonumber\\
		&-\lambda \frac{(W^{h}(t_{1},t_{2}))^{2}}{2N\tau_{2}} p^{v}(t_{1},t_{2}).		
	\end{align}
	\begin{rem}\label{Rem1}
		Note that if we set  $ t=t_{1}-\tau=t_{2}+\tau $, where $ \tau $ is the round trip time in the $ 1D $ model and satisfies the relation $ \tau=\frac{\tau_{1}+\tau_{2}}{2} $, we recover the 1D system.
	\end{rem}
	\subsection{Queue dynamic in two dimensional space}
	
	The queue dynamic relays on the motion of the arrival flow and on the kind of action taken by the AQM (drop or mark). Thus, we provide dynamical equations for each dimension and we separate the cases where ECN is ON or OFF.\\
	\underline{Case 1: ECN ON,}
	\begin{align}
		\frac{\partial q^{h}\left(t_{1}, t_{2}\right)}{\partial t_{1}} = \frac{N}{\tau_{1}(t_{1})} W^{h}(t_{1},t_{2})-C,\\
		\frac{\partial q^{v}\left(t_{1}, t_{2}\right)}{\partial t_{2}} =  \frac{N}{\tau_{1}(t_{1})} W^{v}(t_{1},t_{2})-C.
	\end{align}
	\underline{Case 2: ECN OFF,}
	\begin{align} \label{QueueEcnOffEq8}
		\frac{\partial q^{h}\left(t_{1}, t_{2}\right)}{\partial t_{1}}=&\frac{N}{\tau_{1}}W^{h}(t_{1},t_{2})(1-p^{v}(t_{1},t_{2}-\tau_{2})-C,\\\label{QueueEcnOffEq9}
		\frac{\partial q^{v}\left(t_{1}, t_{2}\right)}{\partial t_{2}}=& \frac{N}{\tau_{2}}W^{v}(t_{1},t_{2})(1-p^{v}(t_{1},t_{2}))-C.
	\end{align}
	
	\begin{rem}
		The idea of modelling the TCP/AQM model in two-dimensional space can be generalized to establish accurate models for interconnected systems, where each subsystem has its own time base and the overall system can be described by means of each dimension. 
	\end{rem}
	
	\begin{rem}
		To perform a stability analysis of the 2D TCP AQM system, we need to solve the 2D differential equations \eqref{Eq2}--\eqref{Eq5}. Since analyzing stability of these nonlinear 2D differential equations is a difficult, if not impossible, task (see the paper \cite{li2022uniform} for uniform stability of 1D nonlinear TCP/AQM model), we are interested in performing a local stability analysis. In this perspective, we look for a tangent system around the equilibrium state. This tangent system, obtained by the first order Taylor expansion, approaches the nonlinear solution under a closed set of initial conditions around the equilibrium point. 
	\end{rem}
	\subsection{Steady state}
	The $ 2D $ nonlinear dynamic model of the TCP/AQM system, \eqref{Eq2}--\eqref{QueueEcnOffEq9}, undergoes approximate linearization around the steady state ($ \hat{W}^{h}, \hat{W}^{v}, \hat{p}, \hat{q}^{h}, \hat{q}^{v} $). The steady state is obtained by solving the equations $ \frac{\partial W^{h}\left(t_{1}, t_{2}\right)}{\partial t_{1}}=0 $, $ \frac{\partial W^{v}\left(t_{1}, t_{2}\right)}{\partial t_{2}}=0 $, $ \frac{\partial q^{h}\left(t_{1}, t_{2}\right)}{\partial t_{1}}=0 $ and $ \frac{\partial q^{v}\left(t_{1}, t_{2}\right)}{\partial t_{2}}=0 $. We obtain,
	\begin{description}
		\item[ECN ON]
		\begin{eqnarray*}
			&&\hat{p}^{v}=\left\lbrace \begin{array} {cc}
				Scenario A & \frac{2N}{2N+\lambda \hat{W}^{h}} \\
				Scenario B & \frac{2N^{2}}{2N^{2}+\lambda (\hat{W}^{h})^{2}}
			\end{array}\right., \\
			&&\left. \begin{array} {cc}
				\hat{W}^{h}=\hat{\tau}_{1}\hat{C},\\
				\hat{W}^{v}=\hat{\tau}_{2}\hat{C},
			\end{array}\right. with \: \begin{array}{c}
				\hat{\tau}_{1}=\frac{\hat{q}^{h}}{\hat{C}}+T_{p}\\
				\hat{\tau}_{2}=\frac{\hat{q}^{v}}{\hat{C}}+T_{p}\\
			\end{array} .
		\end{eqnarray*} 
		\item[ECN OFF]
		\begin{eqnarray*}
			&&\hat{p}^{v}=\left\lbrace \begin{array} {cc}
				Scenario A & \frac{2N}{2N+\lambda \hat{W}^{h}} \\
				Scenario B & \frac{2N^{2}}{2N^{2}+\lambda (\hat{W}^{h})^{2}}
			\end{array}\right., \\
			&&\left. \begin{array} {cc}
				\hat{W}^{h}=\frac{\hat{\tau}_{1}\hat{C}}{1-\hat{p}^{v}},\\
				\hat{W}^{v}=\frac{\hat{\tau}_{2}\hat{C}}{1-\hat{p}^{v}},
			\end{array}\right. with \: \begin{array}{c}
				\hat{\tau}_{1}=\frac{\hat{q}^{h}}{\hat{C}}+T_{p}\\
				\hat{\tau}_{2}=\frac{\hat{q}^{v}}{\hat{C}}+T_{p}\\
			\end{array} .
		\end{eqnarray*} 
		It is worth noted that in steady state $ p^{h} = p^{v} $.
	\end{description}
	
	\subsection{Approximately linearized model}
	The linearization process relies on first-order Taylor series expansion \cite{folland2005higher} and on the related Jacobian matrices. %(See Appendix A).  
	The approximately linearized model of the $ 2D $ linear model is written as a Roesser model \eqref{Eq10},
	\begin{figure*}[!]
		\begin{eqnarray*}
			\hline 
		\end{eqnarray*}
		\begin{eqnarray}\label{Eq10} 
			(\mathcal{S}) :\quad \left\lbrace \begin{array}{cc}
				\left[\begin{array}{c}
					\frac{\partial x^{h}\left(t_{1}, t_{2}\right)}{\partial t_{1}} \\
					\frac{\partial x^{v}\left(t_{1}, t_{2}\right)}{\partial t_{2}}
				\end{array}\right]  =A x\left(t_{1}, t_{2}\right)+A_{\tau} x\left(t_{1}-\tau_{1}, t_{2}-\tau_{2}\right)+Bu(t_{1},t_{2})+B_{\tau}u\left(t_{1}-\tau_{1}, t_{2}-\tau_{2}\right), 
			\end{array} \right.
		\end{eqnarray}
		with, 
		\begin{align*}
			A=&\left[\begin{array}{cc|cc}
				\frac{\delta f_{W}^{h}}{\delta W^{h}} & \frac{\delta f_{W}^{h}}{\delta q^{h}} & \frac{\delta f_{W}^{h}}{\delta W^{v}} & \frac{\delta f_{W}^{h}}{\delta q^{v}} \\
				\frac{\delta f_{q}^{h}}{\delta W^{h}} & \frac{\delta f_{q}^{h}}{\delta q^{h}} & \frac{\delta f_{q}^{h}}{\delta W^{v}} & \frac{\delta f_{q}^{h}}{\delta q^{v}} \\\hline
				\frac{\delta f_{W}^{v}}{\delta W^{h}} & \frac{\delta f_{W}^{v}}{\delta q^{h}} & \frac{\delta f_{W}^{v}}{\delta W^{v}} & \frac{\delta f_{W}^{v}}{\delta q^{v}} \\
				\frac{\delta f_{q}^{v}}{\delta W^{h}} & \frac{\delta f_{q}^{v}}{\delta q^{h}} & \frac{\delta f_{q}^{v}}{\delta W^{v}} & \frac{\delta f_{q}^{v}}{\delta q^{v}}
			\end{array}\right], \quad A_{\tau} = \left[\begin{array}{cc|cc}
				\frac{\delta f_{W}^{h}}{\delta W^{h}_{\tau}} & \frac{\delta f_{W}^{h}}{\delta q^{h}_{\tau}} & \frac{\delta f_{W}^{h}}{\delta W^{v}_{\tau}} & \frac{\delta f_{W}^{h}}{\delta q^{v}_{\tau}} \\
				\frac{\delta f_{q}^{h}}{\delta W^{h}_{\tau}} & \frac{\delta f_{q}^{h}}{\delta q^{h}_{\tau}} & \frac{\delta f_{q}^{h}}{\delta W^{v}_{\tau}} & \frac{\delta f_{q}^{h}}{\delta q^{v}_{\tau}} \\\hline
				\frac{\delta f_{W}^{v}}{\delta W^{h}_{\tau}} & \frac{\delta f_{W}^{v}}{\delta q^{h}_{\tau}} & \frac{\delta f_{W}^{v}}{\delta W^{v}_{\tau}} & \frac{\delta f_{W}^{v}}{\delta q^{v}_{\tau}} \\
				\frac{\delta f_{q}^{v}}{\delta W^{h}_{\tau}} & \frac{\delta f_{q}^{v}}{\delta q^{h}_{\tau}} & \frac{\delta f_{q}^{v}}{\delta W^{v}_{\tau}} & \frac{\delta f_{q}^{v}}{\delta q^{v}_{\tau}}
			\end{array}\right], B=\left[\begin{array}{c|c}
				\frac{\delta f_{W}^{h}}{\delta p^{h}} & \frac{\delta f_{W}^{h}}{\delta p^{v}} \\
				\frac{\delta f_{q}^{h}}{\delta p^{h}} & \frac{\delta f_{q}^{h}}{\delta p^{v}} \\ \hline
				\frac{\delta f_{W}^{v}}{\delta p^{h}} & \frac{\delta f_{W}^{v}}{\delta p^{v}} \\
				\frac{\delta f_{q}^{v}}{\delta p^{h}} & \frac{\delta f_{q}^{v}}{\delta p^{v}}\\
			\end{array}\right], \quad B_{\tau} =\left[\begin{array}{c|c}
				\frac{\delta f_{W}^{h}}{\delta p_{\tau}^{h}} & \frac{\delta f_{W}^{h}}{\delta p_{\tau}^{v}} \\
				\frac{\delta f_{q}^{h}}{\delta p_{\tau}^{h}} & \frac{\delta f_{q}^{h}}{\delta p_{\tau}^{v}} \\ \hline
				\frac{\delta f_{W}^{v}}{\delta p_{\tau}^{h}} & \frac{\delta f_{W}^{v}}{\delta p_{\tau}^{v}} \\
				\frac{\delta f_{q}^{v}}{\delta p_{\tau}^{h}} & \frac{\delta f_{q}^{v}}{\delta p_{\tau}^{v}} \\
			\end{array}\right].
		\end{align*}
		\begin{eqnarray*}
			\hline 
		\end{eqnarray*}
	\end{figure*}
	with the initial conditions,
	\begin{align*}
		x^{h}\left(\theta, t_{2}\right)&=\phi^{h}\left(\theta, t_{2}\right), \quad \forall t_{2} \in \mathbb{R}_{+} \: \mathrm{and} \:  -\tau_{1} \leqslant\theta \leqslant0, \\
		x^{v}\left(t_{1},\theta\right)&=\phi^{v}\left(t_{1}, \theta\right), \quad \forall t_{1} \in \mathbb{R}_{+} \: \mathrm{and} \:-\tau_{2} \leqslant\theta \leqslant0,
	\end{align*}
	where $\phi^{h}$ and $\phi^{v}$ are bounded and have compact support, that is, there exist $L_{1}, L_{2}$, $T_{1}$ and $T_{2}$ such that,
	\begin{align}
		&\left\{\begin{array}{cl}
			\left\|\phi^{h}(\cdot, t)\right\|_{c l} \leqslant L_{1} & \text { if } 0 \leqslant t \leqslant T_{2} \\
			\left\|\phi^{h}(\cdot, t)\right\|_{c l}=0 & \text { if } t>T_{2}
		\end{array}\right.\\
		&\left\{\begin{aligned}
			\left\|\phi^{v}(t, \cdot)\right\|_{c l} \leqslant L_{2} & \text { if } 0 \leqslant t \leqslant T_{1} \\
			\left\|\phi^{v}(t, \cdot)\right\|_{c l}=0 & \text { if } t>T_{1}
		\end{aligned}\right.
	\end{align}
	where,
	\begin{align*}
		&x\left(t_{1}, t_{2}\right)=\left[\begin{array}{c}
			x^{h}\left(t_{1}, t_{2}\right) \\
			x^{v}\left(t_{1}, t_{2}\right)
		\end{array}\right], \\ &x\left(t_{1}-\tau_{1}(t_{1}),t_{2}-\tau_{2}\right)=\left[\begin{array}{c}
			x^{h}\left(t_{1}-\tau_{1}(t_{1}), t_{2}\right) \\
			x^{v}\left(t_{1}, t_{2}-\tau_{2}\right)
		\end{array}\right],
	\end{align*}
	\begin{align*} % etoile signifi la disparition de la numerotation 
		&x^{h}(t_{1},t_{2})=\left[
		\begin{array}{c}
			\delta W^{h}(t_{1},t_{2}) \\
			\delta q^{h}(t_{1},t_{2}) 
		\end{array}
		\right], \quad	x^{v}(t_{1},t_{2})=\left[
		\begin{array}{c}
			\delta W^{v}(t_{1},t_{2}) \\
			\delta q^{v}(t_{1},t_{2}) 
		\end{array}
		\right], \\
		&u(t_{1},t_{2})= \left[\begin{array}{c}
			\delta p^{h}(t_{1},t_{2})\\
			\delta p^{v}(t_{1},t_{2})
		\end{array}\right].
	\end{align*}

	\subsection{Problem definition}
	In this paper, we are interested in designing a stabilizing feedback controller described by, 
	\begin{eqnarray}\label{Controller}
		u(t_{1},t_{2}) &=& Kx(t_{1},t_{2}).
	\end{eqnarray}
	Thus, the $2 \mathrm{D}$ continuous system $ (\mathcal{S}) $ with time  delays is expressed by,
	\begin{align}\label{Eq1}
		(\mathcal{S}) :\quad \left[\begin{array}{c}
			\frac{\partial x^{h}\left(t_{1}, t_{2}\right)}{\partial t_{1}} \\
			\frac{\partial x^{v}\left(t_{1}, t_{2}\right)}{\partial t_{2}}
		\end{array}\right]  =&\mathbb{A} x\left(t_{1}, t_{2}\right) \nonumber\\
	&+\mathbb{A}_{\tau} x\left(t_{1}-\tau_{1}, t_{2}-\tau_{2}\right),
	\end{align}
	where,
	\begin{align*}
		\mathbb{A}=A+BK, \quad \mathbb{A}_{\tau}=A_{\tau}+B_{\tau}K,
	\end{align*}
	and,
	\begin{align*}
		\resizebox{0.9\hsize}{!}{$ 	x\left(t_{1}, t_{2}\right)=\left[\begin{array}{c}
				x^{h}\left(t_{1}, t_{2}\right) \\
				x^{v}\left(t_{1}, t_{2}\right)
			\end{array}\right], \quad x\left(t_{1}-\tau_{1},t_{2}-\tau_{2}\right)=\left[\begin{array}{c}
				x^{h}\left(t_{1}-\tau_{1}, t_{2}\right) \\
				x^{v}\left(t_{1}, t_{2}-\tau_{2}\right)
			\end{array}\right], $}
	\end{align*}
	$x^{h}\left(t_{1}, t_{2}\right) \in \mathbb{R}^{n_{h}}$ is the horizontal state, $x^{v}\left(t_{1}, t_{2}\right) \in \mathbb{R}^{n_{v}}$ is the vertical state, $\tau_{1}$ and $\tau_{2}$ are known constants delays along with horizontal and vertical directions, respectively. 
	
	The main problems considered by this work are defined next, 
	\begin{description}
		\item[\textbf{Problem 1:}] Given positive scalars $ \tau_{1} $ and $ \tau_{2}$, determine analytical conditions for stability assessments of the 0-input 2D system  in \eqref{Eq10}.
		\item[\textbf{Problem 2:}] Given positive scalars $ \tau_{1} $ and $ \tau_{2}$, design feedback gain for stabilizing the 2D system in \eqref{Eq10}.
	\end{description}
	
	The following lemma from \cite{seuret2015hierarchy} will be used in the sequel.
	
	\begin{lem}\label{lemma2} (see \cite{seuret2015hierarchy})
		Let $ x $ be such that $ \dot{x}\in \mathcal{C} $, $ Z\in\mathbb{S}_{n}^{+} $ and $ h>0 $. Then, the inequality 
		\begin{eqnarray}\label{Eq7}
			\resizebox{0.9\hsize}{!}{$ \int_{s}^{b} \dot{x}^{T}(r)Z\dot{x}(r)dr \geq \frac{1}{b-s} \Gamma_{N}^{T}\left[\sum_{k=0}^{N}(2k+1)\pi_{N}^{T}(k)Z\Gamma_{N}(k) \right]\zeta_{N},  $}\nonumber\\
		\end{eqnarray}
		holds, for all integer $ N \in \mathbb{N} $, where,
		\begin{eqnarray*}
			&& \Gamma_{N}= \left\lbrace \begin{array}{cc}
				\left[\begin{array}{cc}
					x^{T}(b) & x^{T}(s)
				\end{array}\right]^{T} & N=0 \\
				\left[\begin{array}{ccccc}
					x^{T}(b) & x^{T}(s) & \frac{1}{\tau} \chi_{0}^{T} & \dots & \frac{1}{\tau} \chi_{N-1}^{T}
				\end{array}\right]^{T} & N>0 \\
			\end{array} \right.  \\
			&&  \pi_{N}(k)=\left\lbrace \begin{array}{cc}
				\left[\begin{array}{cc}
					I & -I
				\end{array}\right] & N=0 \\
				\left[\begin{array}{ccccc}
					I & (-1)^{k+1}I & \theta_{Nk}^{0}I & \dots & \theta_{Nk}^{N-1}I
				\end{array}\right]& N>0 \\
			\end{array} \right. \\
			&&\theta_{Nk}^{j} =\left\lbrace \begin{array}{cc}
				-(2j+1)(1-(-1)^{k+j}) & j \leqslant k \\
				0 & j\geq k
			\end{array}\right., \; \tau= b-s \\
		\end{eqnarray*}	
		with,
		\begin{align*}
			\chi_{i}=\int_{b}^{s}F_{i}(u)x_{t}(u)du, \quad F_{i}(u)=(-1)^{k} \sum_{l=0}^{k} \delta_{l}^{k}\left(\frac{u+h}{h}\right)^{l} ,
		\end{align*}
		and $\delta_{i,l} = (-1)^{l}\left(\begin{array}{c}
			i \\
			l
		\end{array}\right)\left(\begin{array}{c}
			i+l \\
			l
		\end{array}\right)
		$ and $\left(\begin{array}{l}k \\ l\end{array}\right)$ refers to the binomial coefficients given by $\frac{k !}{(k-l) ! l !}$.
	\end{lem}
	
	%\begin{lem}
	%	Operator $ \Delta_{d}: z \rightarrow e_{\Delta} $ satisfies SSG condition $ ||X \circ \Delta_{d} \circ X^{-1}||_{\infty} \leqslant1 $, where $ X $ is a general invertible matrix.
	%\end{lem}
	
	\section{Novel LMI frameworks for stability analysis and feedback gain synthesis of 2D time delay systems}
	The following section presents two main results. The first result permits to  asses the stability of the 0-input system \eqref{Eq10}, whereas the second permits to synthesize a stabilizing feedback gain.
	\subsection{Stability analysis of 2D dynamical systems}
	The result bellow permits to evaluate the stability of the $ 2D $ Roesser models in terms of LMI.
	\begin{thm}\label{THMStability}
		Given positive scalars  $ \tau_{1} $ and $ \tau_{2} $. If there exist positive definite matrices $ P^{v} $, $ P^{h} $, $ Q $, $ R $, and any invertible matrix $ H=diag\left\lbrace H^{h},H^{v}\right\rbrace $ such that the following inequality holds,
		\begin{align}
			\label{THM1}
			\Pi - \Gamma^{T} \left[\begin{array}{cc}
				\mathcal{R}_{1} & 0 \\
				0 &\mathcal{R}_{2}
			\end{array}\right] \Gamma +2M^{T}Hg_{0} \leqslant 0
		\end{align}
		then, the system \eqref{Eq10} with a 0-input is asymptotically stable, where,
		\begin{align}
			e_{i} &= \left[\begin{array}{cc}
				e_{i}^{h} \\
				e_{i}^{v}
			\end{array}\right] = \bordermatrix{&   1   &  \ldots & i-1 & i & i+1 & \ldots & 5 \cr
				& 0 &  \dots  & 0 & I_{n_{h}} &  0  & \dots &  0  \cr
				& 0 &  \dots  & 0 & I_{n_{v}} &  0  & \dots &  0 } 
			\nonumber\\
			&= \bordermatrix{&   1   &  \ldots & i-1 & i & i+1 & \ldots & 5 \cr
				& 0 &  \dots  & 0 & I_{n} &  0  & \dots &  0 }, \nonumber \\
			\label{PiMatrix}\Pi&=  2\mathcal{E}^{hT} \mathcal{A}^{h} +2 \mathcal{E}^{vT}P^{v} \mathcal{A}^{v} +e_{1}^{T}Qe_{1}- e_{2}^{T} Qe_{2} \nonumber\\
			& \quad + e_{5}^{T}\left[\begin{array}{cc}
				\tau_{1}^{2}R^{h} & 0 \\
				0 & \tau_{2}^{2}R^{v}
			\end{array}\right]e_{5} , \\
			g_{0} &= Ae_{1} +A_{\tau}e_{2} -e_{5},  \nonumber \\
			\label{EqEh}\mathcal{E}^{h}&= col\left\lbrace e_{1}^{h},  \tau_{1} e_{3}^{h},  \tau_{1} e_{4}^{h} \right\rbrace ,\\
			\mathcal{E}^{v}&= col\left\lbrace e_{1}^{v},  \tau_{2}  e_{3}^{v},  \tau_{2}  e_{4}^{v} \right\rbrace ,\\
			\mathcal{A}^{h}&=  col\left\lbrace e_{5}^{h},  (e_{1}^{h}-e_{2}^{h}),  (e_{1}^{h}+e_{2}^{h}-2e_{3}^{h}) \right\rbrace ,\\
			\label{EqAv} \mathcal{A}^{v}&=col\left\lbrace e_{5}^{v},  (e_{1}^{v}-e_{2}^{v}),  (e_{1}^{v}+e_{2}^{v}-2e_{3}^{v}) \right\rbrace , \\
			\label{Gamma1} \Gamma&=  col\left\lbrace e_{1}-e_{2}, e_{1}+e_{2}-2e_{4}, e_{1}-e_{2}-6e_{5} \right\rbrace, \\
			\label{MatrixM} M&= e_{1}+e_{2}+e_{5}, \\
			\label{R^{j}} \mathcal{R}&= diag(\mathcal{R}^{h}, \mathcal{R}^{v}) , \quad 	\mathcal{R}^{j}= diag(R^{j},3R^{j}, 5R^{j}), \quad  j=h,v 
		\end{align}
	\end{thm}
	\begin{pf}
		Define the augmented vectors, 
		\begin{align*}
			&\eta^{h}(t_{1},t_{2})= col\left\lbrace x^{h}(t_{1},t_{2}), \tau_{1} \varphi_{1}^{h}(t_{1},t_{2}) \right\rbrace ,  \\
			&\resizebox{1\hsize}{!}{$ \varphi_{1}^{h}(t_{1},t_{2}) =  \frac{1}{\tau_{1}}col\left\lbrace \int_{t_{1}-\tau_{1}}^{t_{1}} x(\alpha,t_{2})d\alpha, \; \int_{t_{1}-\tau_{1}}^{t_{1}} \left\lbrace 2\frac{\alpha-t_{1}+\tau_{1}}{\tau_{1}}-1\right\rbrace x(\alpha,t_{2})d\alpha \right\rbrace  $},
		\end{align*}
		and,
		\begin{align*}
			&\eta^{v}(t_{1},t_{2})= col\left\lbrace x^{v}(t_{1},t_{2}), \tau_{2} \varphi_{1}^{v}(t_{1},t_{2}), \tau_{2}\varphi_{2}^{v}(t_{1},t_{2}) \right\rbrace ,  \\
			&\resizebox{1\hsize}{!}{$\varphi_{1}^{v}(t_{1},t_{2}) = \frac{1}{\tau_{2}}col\left\lbrace \int_{t_{2}-\tau_{2}}^{t_{2}} x(t_{1},\alpha)d\alpha, \; \int_{t_{2}-\tau_{2}}^{t_{2}} \left\lbrace 2\frac{\alpha-t_{2}+\tau_{2}}{\tau_{2}}-1\right\rbrace x(\alpha,t_{2})d\alpha \right\rbrace  $}, 
		\end{align*}
		with, 
		\begin{align*}
			&\resizebox{1\hsize}{!}{$ \left[\begin{array}{cc}
					\varphi_{1}^{h}(t_{1},t_{2})\\
					\varphi_{1}^{v}(t_{1},t_{2})
				\end{array}\right]  = col\left\lbrace \left[\begin{array}{c}
					\frac{1}{\tau_{1}}\int_{t_{1}-\tau_{1}}^{t_{1}} x(\alpha,t_{2})d\alpha\\
					\frac{1}{\tau_{2}} \int_{t_{2}-\tau_{2}}^{t_{2}} x(t_{1},\alpha)d\alpha
				\end{array}\right], \left[\begin{array}{c}
					\frac{1}{\tau_{1}} \int_{t_{1}-\tau_{1}}^{t_{1}} \left\lbrace 2\frac{\alpha-t_{1}+\tau_{1}}{\tau_{1}}-1\right\rbrace x(\alpha,t_{2})d\alpha\\
					\frac{1}{\tau_{2}} \int_{t_{2}-\tau_{2}}^{t_{2}} \left\lbrace 2\frac{\alpha-t_{2}+\tau_{2}}{\tau_{2}}-1\right\rbrace x(\alpha,t_{2})d\alpha
				\end{array}\right] \right\rbrace, $} \\
			&\xi^{T}(t_{1},t_{2})=\left[\begin{array}{cccccccc}
				x(t_{1},t_{2})^{T} & x(t_{1}-\tau_{1},t_{2}-\tau_{2})^{T} & 
			\end{array}\right.\\
			& \quad \left. \begin{array}{cccc}
				\left[\begin{array}{cc}
					\varphi_{1}^{h}(t_{1},t_{2})\\
					\varphi_{1}^{v}(t_{1},t_{2})
				\end{array}\right]^{T} & \left[\begin{array}{c}
					\left(\begin{array}{c}
						\frac{\partial x^{h}\left(t_{1}, t_{2}\right)}{\partial t_{1}} \\
					\end{array}\right)\\
					\left(\begin{array}{c}
						\frac{\partial x^{v}\left(t_{1}, t_{2}\right)}{\partial t_{2}}
					\end{array}\right)
				\end{array}\right]^{T}
			\end{array}\right].
		\end{align*}
		Now we write, $ \left[\begin{array}{c}
			\eta^{h}(t_{1},t_{2})\\
			\eta^{v}(t_{1},t_{2})
		\end{array}\right] $ and $ \left[\begin{array}{c} \dfrac{\partial \eta^{h}(t_{1},t_{2})}{\partial t_{1}} \\
			\dfrac{\partial \eta^{v}(t_{1},t_{2})}{\partial t_{2}} \end{array}\right] $ in term of $ \xi(t_{1},t_{2}) $, we got,
		$
		\left[\begin{array}{c}
			\eta^{h}(t_{1},t_{2})\\
			\eta^{v}(t_{1},t_{2})
		\end{array}\right] = \left[\begin{array}{c}
			\mathcal{E}^{h}\\
			\mathcal{E}^{v}
		\end{array}\right]\xi(t_{1},t_{2}), $ and $ \left[\begin{array}{c} \dfrac{\partial \eta^{h}(t_{1},t_{2})}{\partial t_{1}} \\
			\dfrac{\partial \eta^{v}(t_{1},t_{2})}{\partial t_{2}} \end{array}\right] =\left[\begin{array}{c} \mathcal{A}^{h} \\ \mathcal{A}^{v}\end{array}\right]\xi(t_{1},t_{2}),$
		where $ \mathcal{E}^{h} $, $ \mathcal{E}^{v} $, $ \mathcal{A}^{h} $ and $ \mathcal{A}^{v} $ are defined in \eqref{EqEh} -- \eqref{EqAv}.
		
		Guided by the results in \cite{seuret2015hierarchy}, we first construct a new Lyapunov--Krasovskii functional for the two dimensional time delay systems, as $ V(x(t_{1},t_{2}))=V_{1}(x(t_{1},t_{2}))+V_{2}(x(t_{1},t_{2}))+V_{3}(x(t_{1},t_{2})) $ where,
		\begin{eqnarray}\label{Lyap1}
			V_{1}(x(t_{1},t_{2})) &=& \left[\begin{array}{c}
				\eta^{h}(t_{1},t_{2})\\
				\eta^{v}(t_{1},t_{2})
			\end{array}\right]^{T}P\left[\begin{array}{c}
				\eta^{h}(t_{1},t_{2})\\
				\eta^{v}(t_{1},t_{2})
			\end{array}\right],\\
			V_{2}(x(t_{1},t_{2}))&=&  \int_{t_{1}-\tau_{1}}^{t_{1}}x^{hT}(\theta,t_{2})Q^{h}x^{h}(\theta,t_{2})d\theta \nonumber \\
			&&
			\int_{t_{2}-\tau_{2}}^{t_{2}}x^{vT}(t_{1},\theta)Q^{v}x^{v}(t_{1},\theta)d\theta , \\
			V_{3}(x(t_{1},t_{2})) &=& \resizebox{0.7\hsize}{!}{$\tau_{1}  \int_{-\tau_{1}}^{0} \int_{s}^{t_{1}} \big(\dfrac{\partial x^{h}(\theta,t_{2})}{\partial t_{1}}\big)^{T} R^{h}\big(\dfrac{\partial x^{h}(\theta,t_{2})}{\partial t_{1}}\big)d\theta ds$} \nonumber\\
			&&\label{Lyap3}\resizebox{0.7\hsize}{!}{$ +\tau_{2} \int_{-\tau_{2}}^{0}\int_{s}^{t_{2}} \big(\dfrac{\partial x^{v}(t_{1},\theta)}{\partial t_{2}}\big)^{T}R^{v}\big(\dfrac{\partial x^{v}(t_{1},\theta)}{\partial t_{2}}\big)d\theta ds  $}, \nonumber\\
		\end{eqnarray}
		with,
		\begin{align*}
			\resizebox{1\hsize}{!}{$ 	P = diag\left\lbrace P^{h}, P^{v} \right\rbrace, \quad
				Q = diag\left\lbrace Q^{h}, Q^{v} \right\rbrace, \quad
				R = diag\left\lbrace R^{h}, R^{v} \right\rbrace.  $}
		\end{align*}
		Note that the conditions $ P>0, \; Q>0, \; R>0, $ guarantee the positive definiteness of the Lyapunov--Krasovskii functional. 
		
		Take the divergence operator of the LKF $ V(x(t_{1},t_{2})) $ along the trajectories of system $ (\mathcal{S}) $, yields,
		\begin{align}\label{Lyapunov}
			\operatorname{div} V\left(t_{1}, t_{2}\right)=\frac{\partial V\left(x\left(t_{1}, t_{2}\right)\right)}{\partial t_{1}}+\frac{\partial V\left(x\left(t_{1}, t_{2}\right)\right)}{\partial t_{2}},
		\end{align}
		with,
		\begin{align}\label{Eq26}
			\frac{\partial V\left(x\left(t_{1}, t_{2}\right)\right)}{\partial t_{1}} =& 2\xi^{T}(t_{1},t_{2})\mathcal{E}^{hT}P^{h}\mathcal{A}^{h}\xi(t_{1},t_{2})  \nonumber\\
			& +x^{hT}(t_{1},t_{2})Q^{h}x^{h}(t_{1},t_{2})  \nonumber\\
			& -x^{hT}(t_{1}-\tau_{1},t_{2})Q^{h}x^{h}(t_{1}-\tau_{1},t_{2}) \nonumber \\
			&+\big(\dfrac{\partial x^{h}(t_{1},t_{2})}{\partial t_{1}}\big)^{T}(\tau_{1}^{2}R^{h})\big(\dfrac{\partial x^{h}(t_{1},t_{2})}{\partial t_{1}}\big) \nonumber\\
			&\resizebox{0.7\hsize}{!}{$ - \tau_{1} \int_{t_{1}-\tau_{1}}^{t_{1}} \big(\dfrac{\partial x^{h}(\theta,t_{2})}{\partial t_{1}}\big)^{T} R^{h}\big(\dfrac{\partial x^{h}(\theta,t_{2})}{\partial t_{1}}\big) d\theta $} ,
		\end{align}
		and,
		\begin{align}\label{Eq27}
			\frac{\partial V\left(x\left(t_{1}, t_{2}\right)\right)}{\partial t_{2}} =& 2\xi^{T}(t_{1},t_{2})\mathcal{E}^{vT}P^{v}\mathcal{A}^{v}\xi(t_{1},t_{2})  \nonumber\\
			& +x^{vT}(t_{1},t_{2})Q^{v}x^{v}(t_{1},t_{2}) \nonumber \\
			& +x^{vT}(t_{1},t_{2}-\tau_{2})-Q^{v}x^{v}(t_{1},t_{2}-\tau_{2})  \nonumber \\
			&+\big(\dfrac{\partial x^{v}(t_{1},t_{2})}{\partial t_{2}}\big)^{T}(\tau_{2}^{2}R^{v})\big(\dfrac{\partial x^{v}(t_{1},t_{2})}{\partial t_{2}}\big) \nonumber \\
			&\resizebox{0.7\hsize}{!}{$ -  \tau_{2} \int_{t_{2}-\tau_{2}}^{t_{2}} \big(\dfrac{\partial x^{v}(t_{1},\theta)}{\partial t_{2}}\big)^{T} R^{v}\big(\dfrac{\partial x^{v}(t_{1},\theta)}{\partial t_{2}}\big) d\theta $} .
		\end{align}
		
		Define the non-integral terms of $ \big(\operatorname{div} V(x(t_{1},t_{2})) \big)$ as $ \xi(t_{1},t_{2})\Pi \xi(t_{1},t_{2}) $ where $ \Pi $ is the matrix given in \eqref{PiMatrix}. Then, apply lemma 2 on the integral terms in \eqref{Eq26} and \eqref{Eq27}, by setting $ N=2 $, respectively, will supply,
		\begin{align}\label{Eq28}
			- \tau_{1} \int_{t-\tau_{1}}^{t} \big(\dfrac{\partial x^{h}(\theta,t_{2})}{\partial t_{1}}\big)^{T} R\big(\dfrac{\partial x^{h}(\theta,t_{2})}{\partial t_{1}}\big) d\theta \leqslant \nonumber\\
			- \xi^{T}(t_{1},t_{2})\Gamma^{hT}\mathcal{R}^{h}\Gamma^{h}\xi(t_{1},t_{2}),
		\end{align}
		and, 
		\begin{align}\label{Eq29}
			- \tau_{2} \int_{t-\tau_{2}}^{t} \big(\dfrac{\partial x^{v}(\theta,t_{2})}{\partial t_{1}}\big)^{T} R\big(\dfrac{\partial x^{v}(\theta,t_{2})}{\partial t_{1}}\big) d\theta \leqslant \nonumber\\
			-\xi^{T}(t_{1},t_{2})\Gamma^{vT}\mathcal{R}^{v}\Gamma^{v}\xi(t_{1},t_{2}),
		\end{align}
		where $ \mathcal{R}^{j}, \;j=h,v $ is defined in \eqref{R^{j}}. Note that for given any block diagonal invertible matrix $ H $, we have,
		$$	2\xi^{T}(t_{1},t_{2})\bigg(M^{T}Hg_{0}\bigg)\xi(t_{1},t_{2})=0. $$
		Substitute $ \xi^{T}(t_{1},t_{2})\Pi \xi(t_{1},t_{2}) $, \eqref{Eq28} and \eqref{Eq29} into \eqref{Lyapunov} and summing the results with the above zero equation, supplies,
		\begin{align}\label{Eq30}
			\resizebox{1\hsize}{!}{$ \operatorname{div} V(x(t_{1},t_{2})) \leqslant  \xi^{T}(t_{1},t_{2})\bigg(\Pi - \left[\begin{array}{cc}
					\underset{\Gamma}{\underbrace{\left[\begin{array}{cc}
								\Gamma^{h} & \Gamma^{v}
							\end{array}\right]^{T}}}
				\end{array}\right]^{T} \left[\begin{array}{cc}
					\mathcal{R}_{1} & 0 \\
					0 &\mathcal{R}_{2}
				\end{array}\right] \left[\begin{array}{cc}
					\begin{array}{c}
						\Gamma^{h} \\ \Gamma^{v}
					\end{array}
				\end{array}\right] \bigg) \xi(t_{1},t_{2}) $} \nonumber \\
			+2\xi^{T}(t_{1},t_{2})\bigg(M^{T}Hg_{0}\bigg)\xi(t_{1},t_{2}).
		\end{align}
		If the linear matrix inequalities \eqref{THM1} is satisfied, then, the Lyapunov--Krasovskii functional is decreasing along the two dimensions. $ \square $
	\end{pf}
	\begin{rem}
		It is worth noticing that the LKF in \eqref{Lyap1}--\eqref{Lyap3} is a new  extension of the LKF \cite{seuret2015hierarchy} for 2D time delay systems. To the best of our knowledge, stability analysis of 2D time delay systems has not been addressed with second order Bessel Legendre LKF.
	\end{rem}
	Note that if we apply LMI in the \textit{Theorem \ref{THMStability}} to synthesize the feedback gain, the LMI problem solver will not work because it contains the bilineair terms $HBK$ and $HB_{\tau }K$. The result bellow establishes an equivalent linear form of the result in \textit{Theorem \ref{THMStability}}.
	
	\subsection{Feedback control synthesis}
	The result below permits to synthesis feedback gain matrices to stabilize the control system \eqref{Eq10}.
	
	\begin{thm}
		Given positive scalars  $ \tau_{2} $ and $ \tau_{1} $. If there exist positive definite matrices $ P_{1}^{v} $, $ P_{1}^{h} $, $ Q $, $ Q_{2} $, $ R_{1} $, $ R_{2} $,$ X_{1} $, $ X_{2} $, a matrix gain $ V $ and invertible matrices $ H^{h} $ and $ H^{v} $ such that the following inequality,
		\begin{align}\label{THM2}
			\Pi - \Gamma^{T} \left[\begin{array}{cc}
				\mathcal{R}_{1} & 0 \\
				0 &\mathcal{R}_{2}
			\end{array}\right]\Gamma +2M^{T}g_{1} \leqslant 0
		\end{align}
		holds, then, the closed-loop system \eqref{Eq10} is asymptotically stable where the controller gain is given by $ K=VH^{-1} $ and,
		\begin{align}
			g_{1} &= (AH+BV)e_{1} +(A_{\tau}H+B_{\tau}V)e_{2} -e_{8}.
		\end{align}
	\end{thm}
	
	\begin{pf}
		Consider the closed-loop system \eqref{Eq10}. Apply the change of coordinate $ L(t_{1},t_{2})=H^{-1}x(t_{1},t_{2}) $, hence we have $ x(t_{1},t_{2})=HL(t_{1},t_{2}) $. The partial derivatives of $ L(t_{1},t_{2}) $ along the horizontal and vertical dimensions leads to,
		\begin{eqnarray}\label{Eq33}
			\left[\begin{array}{c}
				\frac{\partial L^{h}\left(t_{1}, t_{2}\right)}{\partial t_{1}} \\
				\frac{\partial L^{v}\left(t_{1}, t_{2}\right)}{\partial t_{2}}
			\end{array}\right]&=&H^{-1}\left[\begin{array}{c}
				\frac{\partial x^{h}\left(t_{1}, t_{2}\right)}{\partial t_{1}} \nonumber \\
				\frac{\partial x^{v}\left(t_{1}, t_{2}\right)}{\partial t_{2}}
			\end{array}\right] \nonumber\\
			&=& H^{-1}(A+BK)HL(t_{1},t_{2}) \nonumber \\
			&&\hspace{-32pt}+ H^{-1}(A_{\tau}+B_{\tau}K)HL(t_{1}-\tau_{1}, t_{2}-\tau_{2}).  
		\end{eqnarray}
		It is clear that the stability of the above dynamical system is equivalent to the stability of \eqref{Eq10}. Consider the variable change $ V=KH $ in \eqref{Eq33}. Then, replace $ \mathbb{A} $, $ \mathbb{A}_{d} $ in the inequality \eqref{THM1} with $ H^{-1}(AH+BV) $, $ H^{-1}(A_{\tau}H+B_{\tau}V) $, respectively. We get \eqref{THM2}. That completes the proof. $ \square $
	\end{pf}
	\begin{rem}
		It is interesting to note that the LMI in \eqref{THM2} is equivalent to the LMI in \eqref{THM1} by using a congruence transformation. For 2-D systems, if the congruence transformation is needed for an LMI linearisation, it is mandatory to apply it with block diagonal matrices characterizing the horizontal and vertical dimensions, see e.g. \cite{el2013robust}. In a similar way, when the change of coordinates is used, it is mandatory to consider a block diagonal matrix like in our case the matrix $H$ is set $ diag\{H^{h},H^{v}\} $.
	\end{rem}

	\section{Simulations}
	The present section gives the simulation results for some different scenarios and concludes based on detailed comparison on the performance of the proposed protocol in Internet. 
	
	\begin{figure}[htb]
		\centering
		\includegraphics[width=1\linewidth]{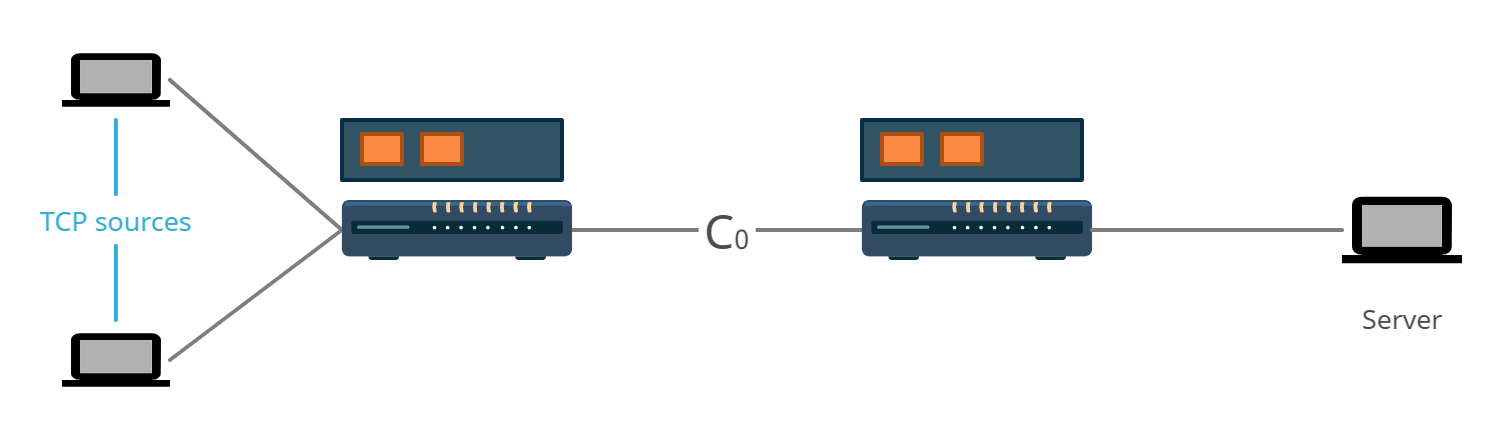}
		\caption{Dumbbell topology.}
		\label{fig:dumbbelltopology}
	\end{figure}
	
	Consider the topology of the network depicted in Fig. \ref{fig:dumbbelltopology}, governed by $ 200 $ loads for scenario B, and $ 800 $ loads for scenario A, and the link bandwidth $ C_{0}=10 Mbit/s $. The queue reference $ Q_{ref}=2000 \; packets $, and a propagation delay $ T_{p}=0.001 $. For scenario A, the average congestion window in the horizontal and vertical dimensions are $ \hat{W}^{v} = \hat{W}^{h} = 1.3282 $ with $ 800 $ loads, $ \hat{p}^{v}=\hat{p}^{h}=0.6001 $ and $ \lambda=1 $. For scenario B, the average congestion window in the horizontal and vertical dimensions are $ \hat{W}^{v} = \hat{W}^{h} = 5.3128 $, with $ 200 $ loads, $ \hat{p}^{v}=\hat{p}^{h}=0.0662 $ and $ \lambda=2.9450 $. 
	%\FloatBarrier
	
	For the 0-input system, we find that the inequality \eqref{THM1} is infeasible. We checked if the system was stable by simulation and found it unstable (figures for this simulation are not provided). Thus, we go into our analysis and solve \eqref{THM2} with the feedback controller. We find the inequality \eqref{THM2} feasible. For replicability, the state space matrices and the gain matrices $ K_{j} $, with the $ j=A,B $ denotes the scenario, are given by,
	\begin{itemize}
		\item[$ \bullet $ ] Scenario A, we obtain:
		\begin{align*}
			&	\resizebox{0.9\hsize}{!}{$ A= \left[\begin{array}{cccc}
					-0.0024941 & -1.6563e-06 & 0      & 0      \\
					1596.4     & 1.0602      & 0      & 0      \\
					1.9831     & 0           & 0      & 0      \\
					0          & 0           & 1588.5 & 1.0496
				\end{array}\right], \; B= \left[\begin{array}{cc}
					0      &      0\\
					-5312.8   &         0\\
					0     & -6.6465\\
					0     & -5286.4
				\end{array}\right] $}\\
			& \resizebox{0.9\hsize}{!}{$ A_{\tau}= \left[\begin{array}{cccc}
					1.993 & 0.0013252 & 0 & 0         \\
					0     & 0         & 0 & 0         \\
					0     & 0         & 0 & 0.0013104 \\
					0     & 0         & 0 & 0
				\end{array}\right], \;  B_{\tau}= \left[\begin{array}{cc}
					0   &   -6.6465 \\
					0    &        0\\
					0     &       0\\
					0      &      0
				\end{array}\right] $} \\
			&K_{A}= \left[\begin{array}{cccc}
				1.409040145867033  & 0.001050302085212  \\
				0.261050119747904  & 0.000106642597102  
			\end{array}\right.\\
			&\left.\begin{array}{cc}
				-0.442659495007931 & -0.000665182278239 \\
				0.377269799513580  & 0.000536355427763
			\end{array}\right]
		\end{align*}
		\item[$ \bullet $ ] Scenario B, we obtain:
		\begin{align*}
			&	\resizebox{0.9\hsize}{!}{$ A= \left[\begin{array}{cccc}
					-175.8330 & 0       & 0        & 0      \\
					933.8000  & 2.4805  & 0        & 0      \\
					929.1285  & -2.4564 & 0        & 2.4554 \\
					0         & 0       & 929.1542 & 2.4559
				\end{array}\right], \; B= 10^{3} \times \left[\begin{array}{cc}
					0      &   0 \\
					-5.3128  &       0 \\
					0  & -5.2874 \\
					0  & -5.2864
				\end{array}\right] $}\\
			& \resizebox{0.9\hsize}{!}{$ A_{\tau}= \left[\begin{array}{cccc}
					175.7512  &  0.4669 &        0    &     0\\
					0      &   0    &     0      &   0\\
					0      &   0   &      0  &  0.0122\\
					0      &   0      &   0   &      0
				\end{array}\right], \;  B_{\tau}= 10^{3}\times\left[\begin{array}{cc}
					0   &-1.0010 \\
					0    &     0\\
					0     &    0\\
					0      &   0
				\end{array}\right] $} \\
			&	K_{B}= \left[\begin{array}{cccc}
				0.194800805211746 & 0.067127479835209 \\
				0.167303800511358 & -0.000437270587770
			\end{array}\right.\\
			&	\left.\begin{array}{cc}
				-0.000657956320659 & -0.000033555865312\\
				0.017422461138873  & 0.000880352801977
			\end{array}\right]
		\end{align*}
	\end{itemize}
	The initial condition is set,
	\begin{eqnarray}
		x_{0}^{T}&=& \left[\begin{array}{cccc}
			-1 & -20 & -1 & -20
		\end{array}\right],
	\end{eqnarray}
	
	\begin{figure}[!h]
		\begin{minipage}[t]{0.48\linewidth}
			\centering
			\includegraphics[width=1\linewidth]{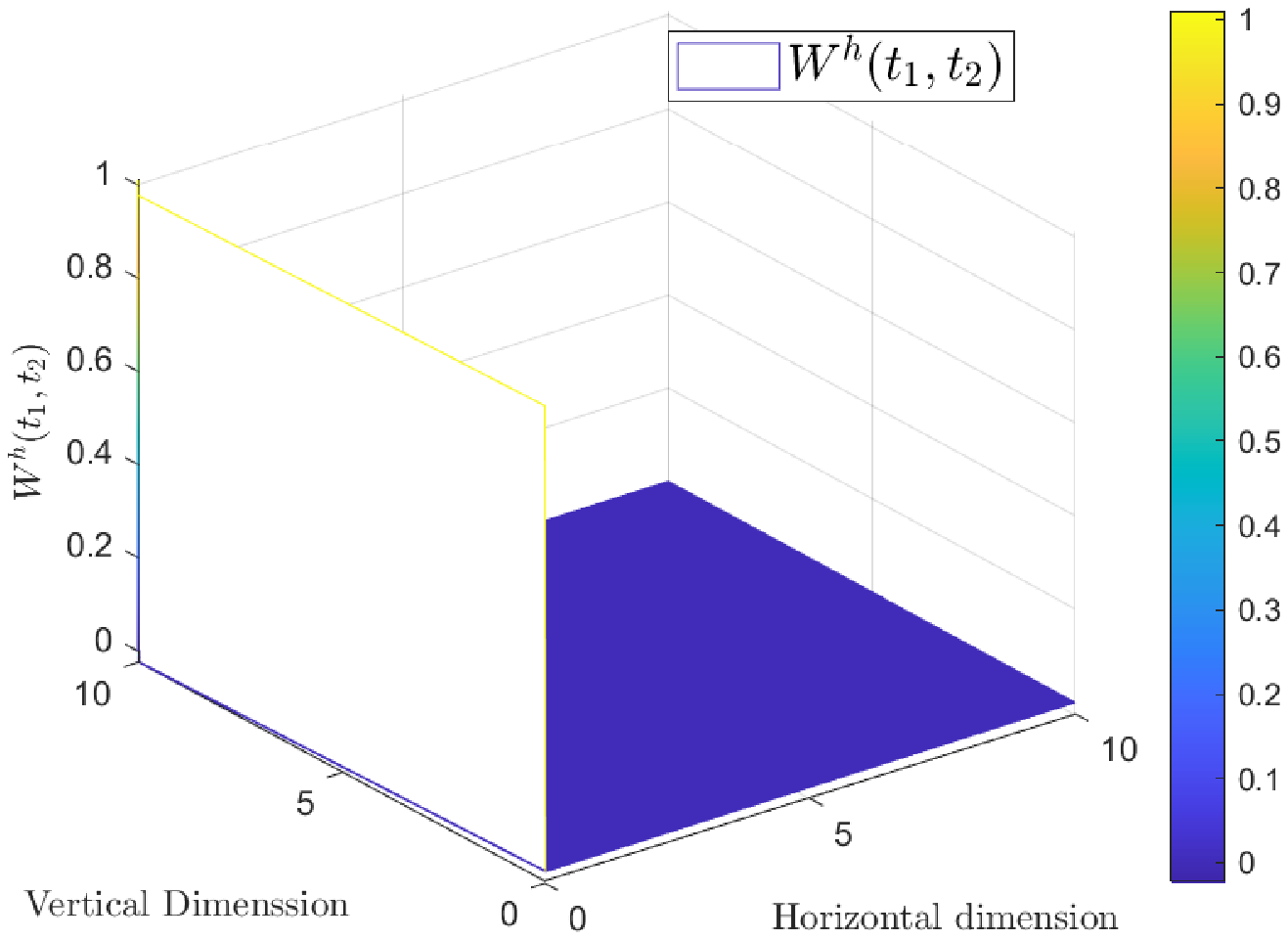}
		\end{minipage}
		\hspace{0.1cm}
		\begin{minipage}[t]{0.48\linewidth} 
			\centering
			\includegraphics[width=1\linewidth]{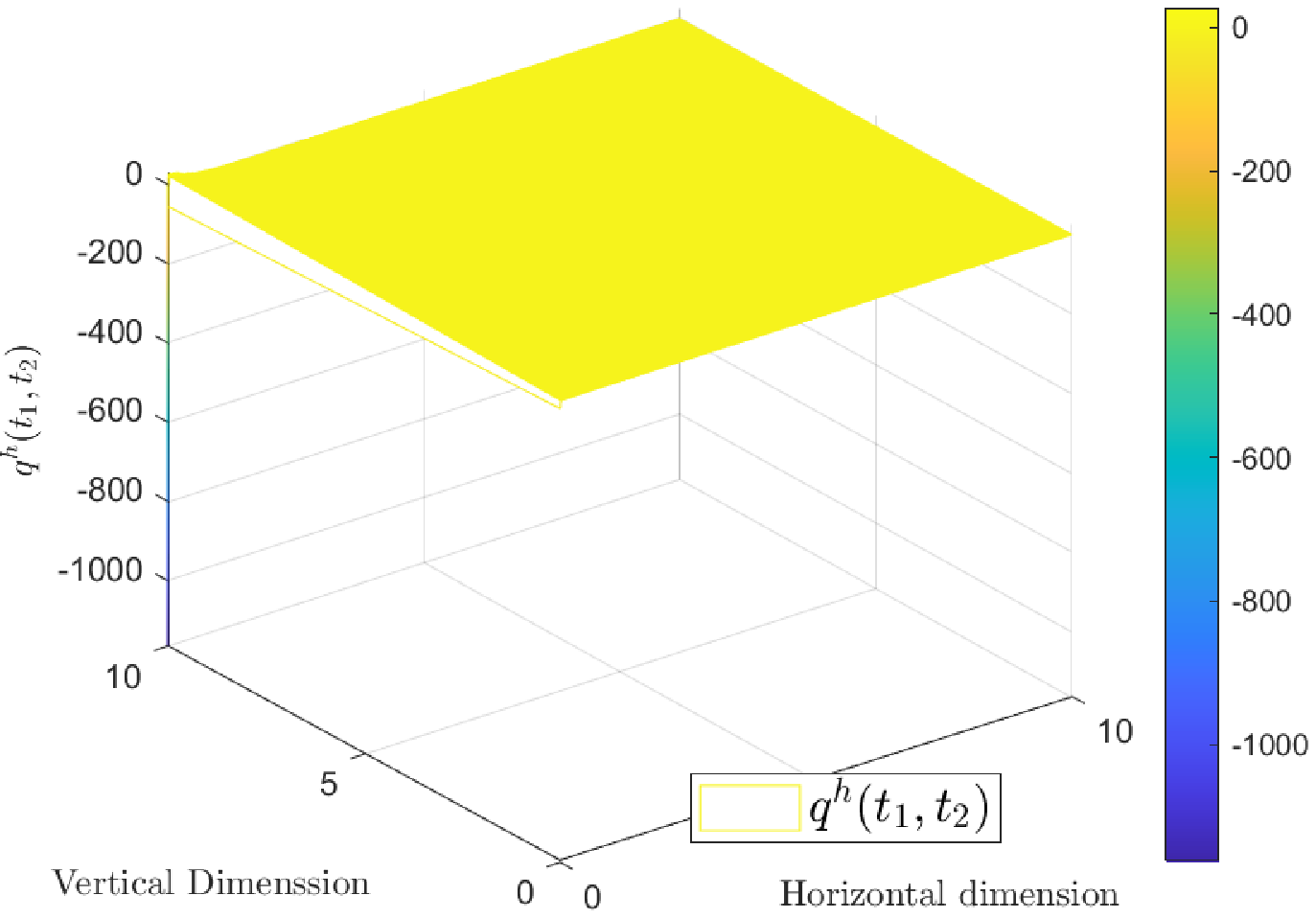}
		\end{minipage}        
		\caption{Scenario A, horizontal state}
		\label{fig:dumbbellscenaHorizontalA}
	\end{figure}
	
	\begin{figure}[!h]
		\begin{minipage}[t]{0.48\linewidth}
			\centering
			\includegraphics[width=1\linewidth]{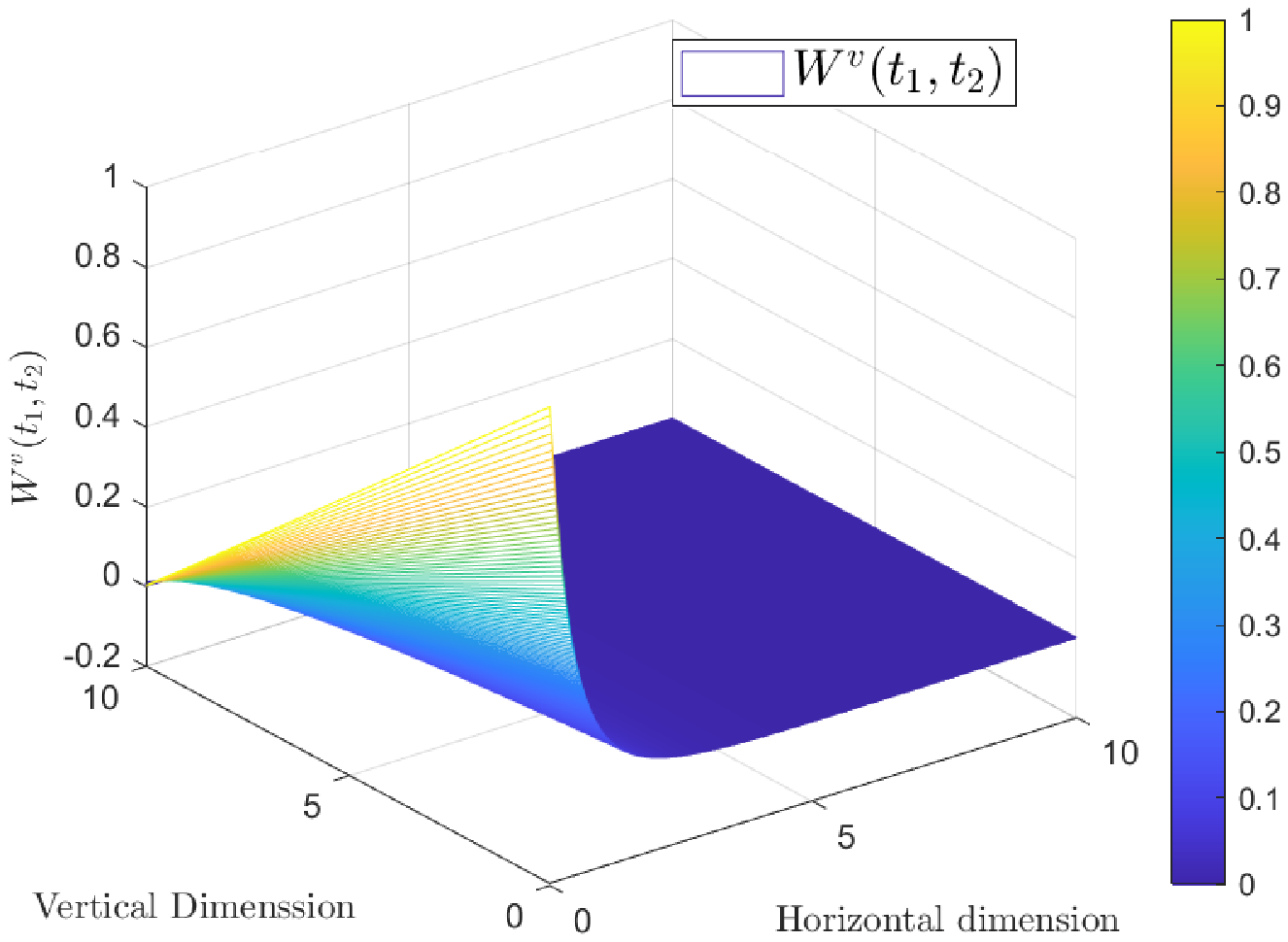}
		\end{minipage}
		\hspace{0.1cm}
		\begin{minipage}[t]{0.48\linewidth} 
			\centering
			\includegraphics[width=1\linewidth]{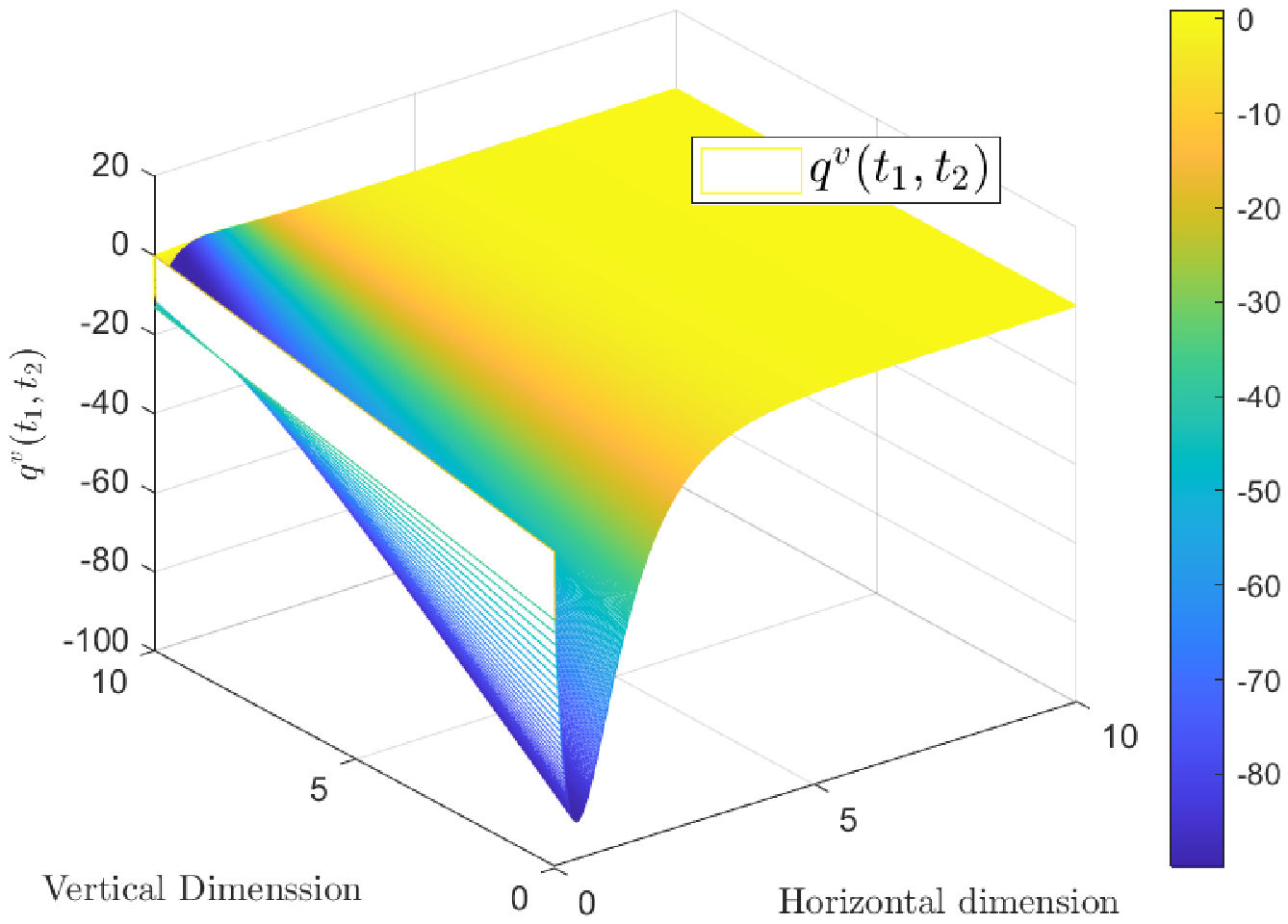}
		\end{minipage}
		\caption{Scenario A, vertical state.}
		\label{fig:dumbbellscenaVerticalA}        
	\end{figure}

	\begin{figure}[!h]
		\begin{minipage}[t]{0.48\linewidth}
			\centering
			\includegraphics[width=1\linewidth]{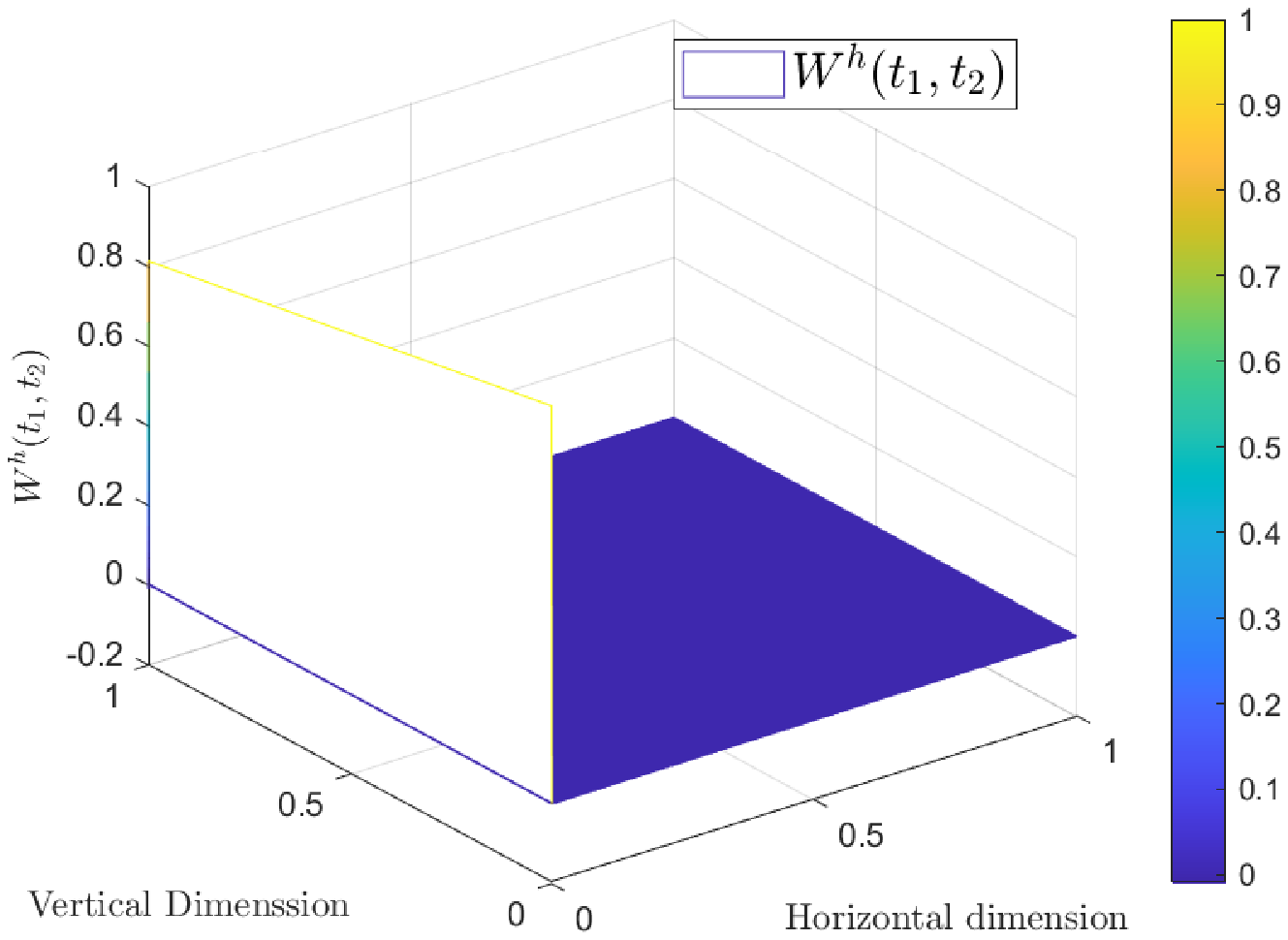}
		\end{minipage}
		\hspace{0.1cm}
		\begin{minipage}[t]{0.48\linewidth} 
			\centering
			\includegraphics[width=1\linewidth]{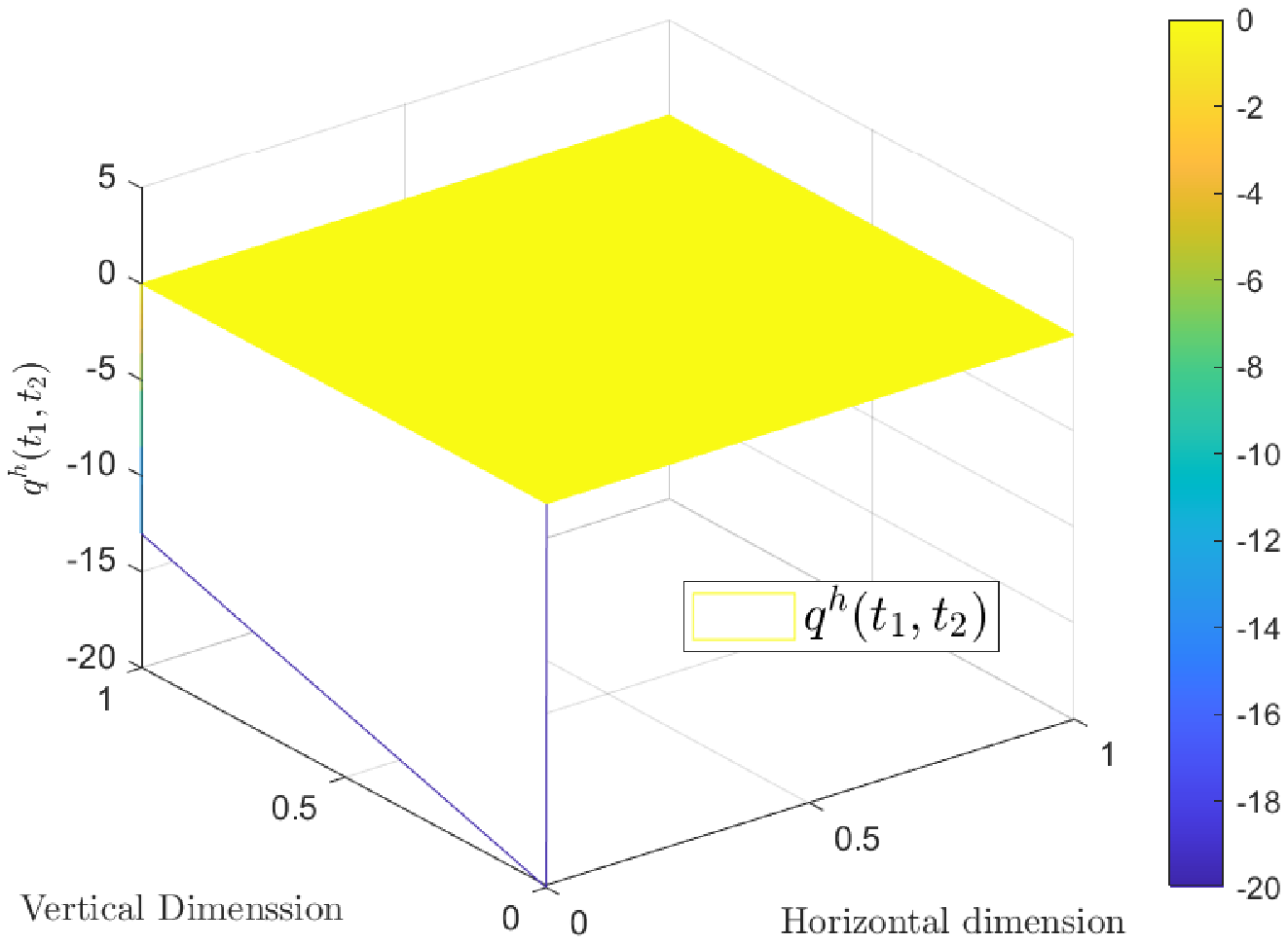}
		\end{minipage}        
		\caption{Scenario B, horizontal state}
		\label{fig:dumbbellscenaHorizontalB}
	\end{figure}
	
	\begin{figure}[!h]
		\begin{minipage}[t]{0.48\linewidth}
			\centering
			\includegraphics[width=1\linewidth]{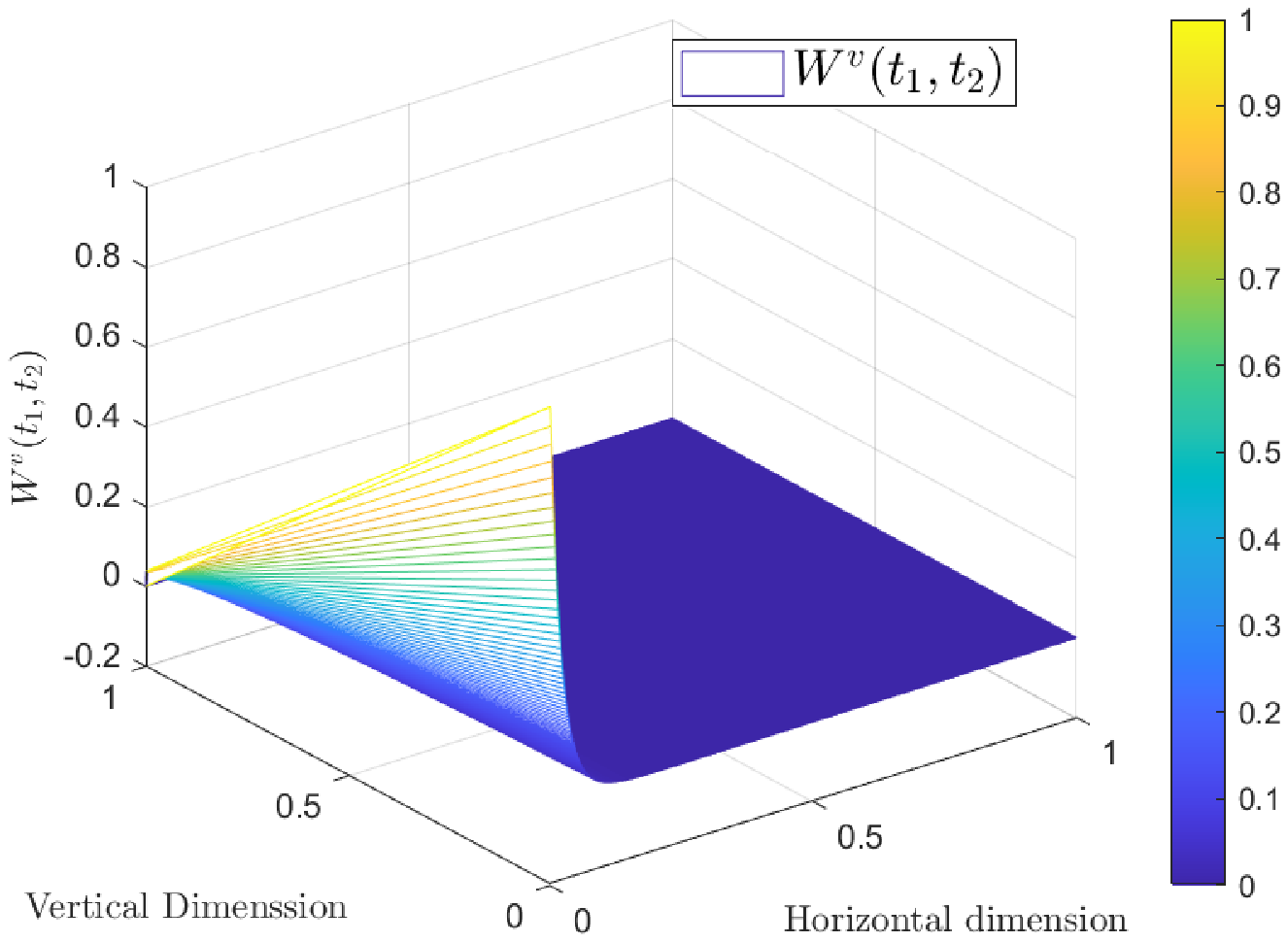}
		\end{minipage}
		\hspace{0.1cm}
		\begin{minipage}[t]{0.48\linewidth} 
			\centering
			\includegraphics[width=1\linewidth]{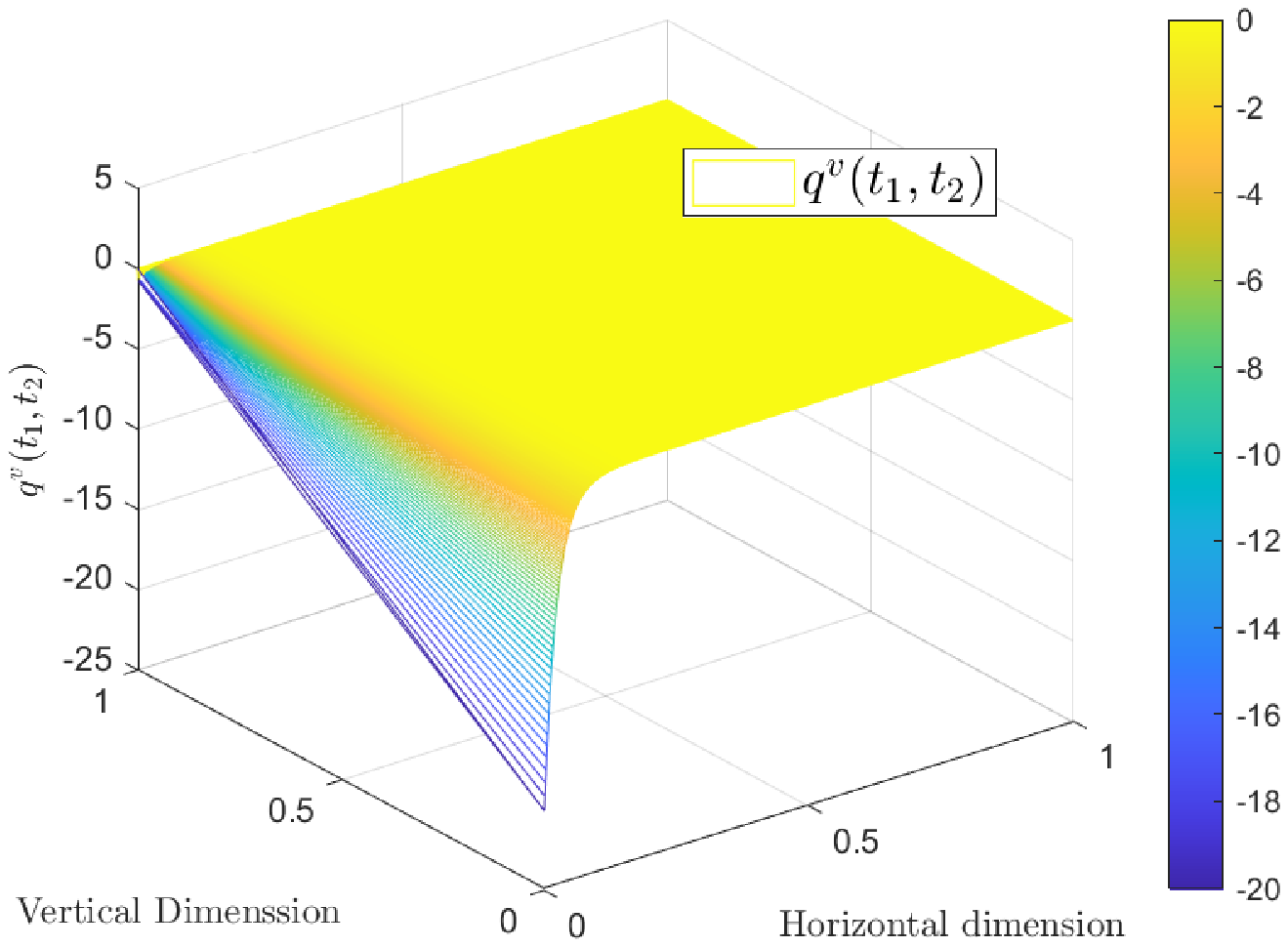}
		\end{minipage}
		\caption{Scenario B, vertical state.}
		\label{fig:dumbbellscenaVerticalB}        
	\end{figure}
	
	%\FloatBarrier
	Figs. \ref{fig:dumbbellscenaHorizontalA} -- \ref{fig:dumbbellscenaVerticalB} show the dynamic evolution of the TCP and router queue for scenarios A and B along both dimensions. The stability of the queue is marked in yellow while that of the congestion window is marked in blue. From the queue responses, we can see, especially for the vertical dynamics, that it is first stabilized at horizontal time and then at vertical time, with a small latency in between. This leads to the conclusion that, although the AQM acts on the vertical dimension, it first affects the horizontal dimension, which is the time base with respect to the sender. On the side of figures \ref{fig:dumbbellscenaVerticalA} and \ref{fig:dumbbellscenaVerticalB}, they show that the stabilization of the congestion window takes place on the vertical dimension and then on the horizontal dimension. This view from the two sides will allow to make new remarks about the operation of TCP in response to an AQM, which was previously limited to a single time base. Second, it is now clearer how the actions of an AQM affect the quality of service of the Internet with respect to what happens at the router and what happens at the sender's congestion window in the same time.

	\section{Conclusion}
	
	The main contribution of this work is a novel and versatile fluid model for TCP/AQM analysis. The model is designed from two temporal basis points of view, resulting into two dimensional differential equations. This framework offers opportunities to analyze the stability of a diverse set of controllers that could be linear or nonlinear in one dimensional or two dimensional spaces. First, we show that the new model is more general and can be reduced to the one dimensional model in \cite{xu2015new}, (see remark \ref{Rem1}). We go on to analyze our 2D fluid model and found that the system has a unique equilibrium point. We then deduce a linear approximation of the model using the first order Taylor expansion. At this stage, for the first time a two dimensional second order Bessel Legendre Lyapunov functional is constructed. This latter permits to derive one LMI condition for stability assessment of 2D time delay systems and a second one for feedback gain synthesis. Finally, some simulations using MATLAB show that the feedback controller achieves a stabilisation of the 2D system's solution.
	
	In conclusion, according to the resulting analytical LMI condition, the considered two traffic scenarios with a 0-input are unstable. Our second conclusion is that the coordinate change, or any equivalent transformation such as the congruence transformation, used for 2D systems requires a diagonal block matrix where the first block is characterized by the horizontal dimension and the second by the vertical dimension.
	
	As a future research line, we are planning to discuss the feedback implementation constraint of the 2D controller and validate the controlled model using network simulator 3 (NS3).

	%	\begin{ack}
	%		A.J. Rojas is grateful for the support of ANID, through Basal Project AC3E grant FB0008.
	%	\end{ack}
	
	\bibliography{ifacconf}             % bib file to produce the bibliography

	\appendix
	% in the appendices.
	
	\section*{Appendix A. Linearisation}
	The First-order Taylor Series expansion, \cite{folland2005higher}, of the non-linear delay differential equations \eqref{Eq2}--\eqref{QueueEcnOffEq9} are as follows,
	\begin{eqnarray}\label{A.1}
		\frac{ \partial	\delta W^{h}(t_{1},t_{2})}{\partial t_{1} }\hspace{-3pt}&=& \hspace{-2pt} \frac{\delta f_{W}^{h}}{\delta W^{h}} \delta W^{h}(t_{1},t_{2}) \hspace{-3pt}+\hspace{-3pt} \frac{\delta f_{W}^{h}}{\delta W_{\tau}^{h}} \delta W^{h}(t_{1}-\tau(t_{1}),t_{2}) \nonumber\\
		&&\hspace{-15pt}+\hspace{-1pt}\frac{\delta f_{W}^{h}}{\delta W^{v}} \delta W^{v}(t_{1},t_{2}) \hspace{-1pt}+\hspace{-1pt} \frac{\delta f_{W}^{h}}{\delta W_{\tau}^{v}} \delta W^{v}(t_{1},t_{2}-\tau(t_{2})) \nonumber\\
		&&\hspace{-1pt}+\hspace{-1pt} \frac{\delta f_{W}^{h}}{\delta p} \delta p^{h}(t_{1},t_{2}) \hspace{-1pt}+\hspace{-1pt} 	\frac{\delta f_{W}^{h}}{\delta p_{\tau}} \delta p^{h}(t_{1}-\tau(t_{1}),t_{2})  \nonumber\\
		&&\hspace{-1pt}+\hspace{-1pt}\frac{\delta f_{W}^{h}}{\delta p} \delta p^{v}(t_{1},t_{2}) \hspace{-1pt}+\hspace{-1pt} 	\frac{\delta f_{W}^{h}}{\delta p_{\tau}} \delta p^{v}(t_{1},t_{2}-\tau(t_{2})) \nonumber \\
		&& \hspace{-1pt}+\hspace{-1pt} \frac{\delta f_{W}^{h}}{\delta q^{h}} \delta q^{h}(t_{1},t_{2})  \hspace{-1pt}+\hspace{-1pt} \frac{\delta f_{W}^{h}}{\delta q_{\tau}^{h}} \delta q^{h}(t_{1}-\tau(t_{1}),t_{2}) \nonumber\\
		&&\hspace{-1pt}+\hspace{-1pt} \frac{\delta f_{W}^{h}}{\delta q^{v}} \delta q^{v}(t_{1},t_{2})  \hspace{-1pt}+\hspace{-1pt} \frac{\delta f_{W}^{h}}{\delta q_{\tau}^{v}} \delta q^{v}(t_{1},t_{2}-\tau(t_{2}))\nonumber\\
		&& \hspace{-1pt}+\hspace{-1pt} h.o.t,  \nonumber\\
		\frac{ \partial	\delta W^{v}(t_{1},t_{2})}{\partial t_{2} }\hspace{-3pt}&=& \hspace{-2pt} \frac{\delta f_{W}^{v}}{\delta W^{h}} \delta W^{h}(t_{1},t_{2}) \hspace{-3pt}+\hspace{-3pt} \frac{\delta f_{W}^{v}}{\delta W_{\tau}^{h}} \delta W^{h}(t_{1}-\tau(t_{1}),t_{2}) \nonumber\\
		&&\hspace{-15pt}+\hspace{-1pt}\frac{\delta f_{W}^{v}}{\delta W^{v}} \delta W^{v}(t_{1},t_{2}) \hspace{-1pt}+\hspace{-1pt} \frac{\delta f_{W}^{v}}{\delta W_{\tau}^{v}} \delta W^{v}(t_{1},t_{2}-\tau(t_{2})) \nonumber\\
		&&\hspace{-1pt}+\hspace{-1pt} \frac{\delta f_{W}^{v}}{\delta p} \delta p^{h}(t_{1},t_{2}) \hspace{-1pt}+\hspace{-1pt} 	\frac{\delta f_{W}^{v}}{\delta p_{\tau}} \delta p^{h}(t_{1}-\tau(t_{1}),t_{2}) \nonumber\\
		&& \hspace{-1pt}+\hspace{-1pt}\frac{\delta f_{W}^{v}}{\delta p} \delta p^{v}(t_{1},t_{2}) \hspace{-1pt}+\hspace{-1pt} 	\frac{\delta f_{W}^{v}}{\delta p_{\tau}} \delta p^{v}(t_{1},t_{2}-\tau(t_{2})) \nonumber \\
		&& \hspace{-1pt}+\hspace{-1pt} \frac{\delta f_{W}^{v}}{\delta q^{h}} \delta q^{h}(t_{1},t_{2})  \hspace{-1pt}+\hspace{-1pt} \frac{\delta f_{W}^{v}}{\delta q_{\tau}^{h}} \delta q^{h}(t_{1}-\tau(t_{1}),t_{2}) \nonumber\\
		&&\hspace{-1pt}+\hspace{-1pt} \frac{\delta f_{W}^{v}}{\delta q^{v}} \delta q^{v}(t_{1},t_{2})  \hspace{-1pt}+\hspace{-1pt} \frac{\delta f^{v}}{\delta q_{\tau}^{v}} \delta q^{v}(t_{1},t_{2}-\tau(t_{2})) \nonumber\\
		&&\hspace{-1pt}+\hspace{-1pt} h.o.t, 
	\end{eqnarray}
	where $ h.o.t. $ denotes higher-order terms, and other notations are as follow, $ f_{W}^{h} := \frac{\partial W^{h}\left(t_{1}, t_{2}\right)}{\partial t_{1}} $, $ f_{W}^{v} := \frac{\partial W^{v}\left(t_{1}, t_{2}\right)}{\partial t_{2}} $, $  W_{\tau}^{h}(t_{1},t_{2})= W^{h}(t_{1}-\tau(t_{1}),t_{2}),  q_{\tau}^{h}(t_{1},t_{2})= q^{h}(t_{1}-\tau(t_{1}),t_{2}),  p_{\tau}^{h}(t_{1},t_{2})= p^{h}(t_{1}-\tau(t_{1}),t_{2})  $, $  W_{\tau}^{v}(t_{1},t_{2})= W^{v}(t_{1}),t_{2}-\tau(t_{2}),  q_{\tau}^{v}(t_{1},t_{2})= q^{v}(t_{1},t_{2}-\tau(t_{2}),  p_{\tau}^{v}(t_{1},t_{2})= p^{v}(t_{1},t_{2}-\tau(t_{2}))  $ and $ \delta W^{j}(t_{1},t_{2}) =  W^{j}(t_{1},t_{2}) - W_{0}^{j} , $  $ \delta q(t_{1},t_{2}) = q^{j}(t_{1},t_{2}) - q_{0}^{j}, \delta p^{j}(t_{1},t_{2}) = p^{j}(t_{1},t_{2}) - p_{0}^{j} $, $ \quad j=h,v$. The partial derivatives for scenarios A and B respectively are given by,
	
	\begin{description}
		\item[Scenario A: ] Horizontal dimension
		\begin{align*}
			\hspace{-10pt}&\resizebox{1\hsize}{!}{$ \frac{\delta f_{W}^{h}}{\delta W^{h}}= -\frac{\lambda \hat{W}^{h} \hat{p}^{v}}{2N\hat{\tau}_{1}},\frac{\delta f_{W}^{h}}{\delta q_{\tau}^{h}} =  \frac{\hat{W}^{h}(1-\hat{p}^{v})}{\hat{\tau}_{1}^{2}C} ,
				\frac{\delta f_{W}^{h}}{\delta W_{\tau}^{h}} = \frac{1-\hat{p}^{v}}{\tau_{1}}-\frac{\lambda \hat{W} \hat{p}^{v}}{2N\hat{\tau}_{1}}, $}  \\
			\hspace{-10pt}&\resizebox{1\hsize}{!}{$\frac{\delta f_{W}^{h}}{\delta W^{v}} = 0,
			\frac{\delta f_{W}^{h}}{\delta W_{\tau}^{v}} = 0, 
			\frac{\delta f_{W}^{h}}{\delta p^{h}}= 0 , \frac{\delta f_{W}^{h}}{\delta p_{\tau}^{h}} = 0,  \frac{\delta f_{W}^{h}}{\delta p^{v}} = 0 ,  \frac{\delta f_{W}^{h}}{\delta q^{v}} =0,$}\\
			\hspace{-10pt}& \resizebox{1\hsize}{!}{$\frac{\delta f_{W}^{h}}{\delta p_{\tau}^{v}} = -\frac{\hat{W}^{h}}{\hat{\tau}_{1}} - \frac{\lambda (\hat{W}^{h})^{2}}{2N\hat{\tau}_{1}} ,\frac{\delta f_{W}^{h}}{\delta q^{h}} =- \frac{\lambda  (\hat{W}^{h})^{2}\hat{p}^{v}}{2N\hat{\tau}_{1}^{2}C},  \frac{\delta f_{W}^{h}}{\delta q_{\tau}^{v}} = 0, $}
		\end{align*}
		\item[Scenario A: ] Vertical dimension
		\begin{align*}
			\hspace{-10pt}&\resizebox{1\hsize}{!}{$\frac{\delta f_{W}^{v}}{\delta W^{h}} = \frac{(1-\hat{p}^{v})}{\hat{\tau}_{2}} - \frac{\lambda \hat{W}^{h}\hat{p}^{v}}{N\hat{\tau}_{2}},
			\frac{\delta f_{W}^{v}}{\delta W_{\tau}^{h}} =0,
			\frac{\delta f_{W}^{v}}{\delta W^{v}} = 0,
			\frac{\delta f_{W}^{v}}{\delta W_{\tau}^{v}} = 0,$} \\
			\hspace{-10pt}&\frac{\delta f_{W}^{v}}{\delta p^{h}} = 0 , \frac{\delta f_{W}^{v}}{\delta p_{\tau}^{h}} = 0,  \frac{\delta f_{W}^{v}}{\delta p^{v}} = \frac{-\hat{W}^{h}}{\hat{\tau_{2}}} - \frac{\lambda (\hat{W}^{h})^{2}}{2N\hat{\tau}_{2}} , \frac{\delta f_{W}^{v}}{\delta p_{\tau}^{v}} =0 ,\\
			\hspace{-10pt}&\resizebox{1\hsize}{!}{$ \frac{\delta f_{W}^{v}}{\delta q^{h}} =0,  \frac{\delta f_{W}^{v}}{\delta q_{\tau}^{h}} =  0 , \frac{\delta f_{W}^{v}}{\delta q^{v}} = \frac{\hat{W}^{h}(1-p^{v})}{\hat{\tau}_{2}^{2}C}- \frac{\lambda  (\hat{W}^{h})^{2}\hat{p}^{v}}{2N\hat{\tau}_{2}^{2}C},  \frac{\delta f_{W}^{v}}{\delta q_{\tau}^{v}} = 0. $}
		\end{align*}
		\item[Scenario B: ] Horizontal dimension
		\begin{align*}
			\hspace{-10pt}&\resizebox{1\hsize}{!}{$\frac{\delta f_{W}^{h}}{\delta W^{h}} = -\frac{N(1-p^{v})}{\hat{\tau}_{1}\hat{W}^{h}}- \frac{\lambda (\hat{W}^{h})^{2} \hat{p}^{v}}{2N\hat{\tau}_{1}},
			\frac{\delta f_{W}^{h}}{\delta W_{\tau}^{h}} = \frac{N(1-\hat{p}^{v})}{\tau_{1}\hat{W}^{h}}-\frac{\lambda \hat{W}^{h} \hat{p}^{v}}{2N\hat{\tau}_{1}}, $}\\
		\hspace{-10pt}&\resizebox{1\hsize}{!}{$ \frac{\delta f_{W}^{h}}{\delta p^{h}} = 0 , \frac{\delta f_{W}^{h}}{\delta p_{\tau}^{h}} = 0,  \frac{\delta f_{W}^{h}}{\delta p^{v}} = 0 , \frac{\delta f_{W}^{h}}{\delta p_{\tau}^{v}} = -\frac{N}{\hat{\tau}_{1}} - \frac{\lambda (\hat{W}^{h})^{2}}{2N\hat{\tau}_{1}} , 
			\frac{\delta f_{W}^{h}}{\delta W_{\tau}^{v}} = 0, $}  \\
		\end{align*}
		\begin{align*}
			\hspace{-10pt}&\resizebox{1\hsize}{!}{$ \frac{\delta f_{W}^{h}}{\delta q^{h}} =0,  \frac{\delta f_{W}^{h}}{\delta q_{\tau}^{h}} =  \frac{N(1-\hat{p}^{v})}{\hat{\tau}_{1}^{2}C} , \frac{\delta f_{W}^{h}}{\delta q^{v}} =0,  \frac{\delta f_{W}^{h}}{\delta q_{\tau}^{v}} = 0,
				\frac{\delta f_{W}^{h}}{\delta W^{v}} = 0, $}\\
		\end{align*}
		\item[Scenario B: ] Vertical dimension
		\begin{align*}
			\hspace{-10pt}&\resizebox{1\hsize}{!}{$\frac{\delta f_{W}^{v}}{\delta W^{h}} = \frac{N(1-\hat{p}^{v})}{\hat{\tau}_{2}} - \frac{\lambda \hat{W}^{h}\hat{p}^{v}}{N\hat{\tau}_{2}},
			\frac{\delta f_{W}^{v}}{\delta W_{\tau}^{h}} =0,
			\frac{\delta f_{W}^{v}}{\delta W^{v}} = 0,
			\frac{\delta f_{W}^{v}}{\delta W_{\tau}^{v}} = 0,$} \\
			\hspace{-10pt}&\resizebox{1\hsize}{!}{$\frac{\delta f_{W}^{v}}{\delta p^{h}} = 0 , \frac{\delta f_{W}^{v}}{\delta p_{\tau}^{h}} = 0,  \frac{\delta f_{W}^{v}}{\delta p^{v}} = \frac{-N\hat{W}^{h}}{\hat{\tau_{2}}} - \frac{\lambda (\hat{W}^{h})^{2}}{2N\hat{\tau}_{2}} , \frac{\delta f_{W}^{v}}{\delta p_{\tau}^{v}} =0 ,$}\\
			\hspace{-10pt}&\resizebox{1\hsize}{!}{$\frac{\delta f_{W}^{v}}{\delta q^{h}} =0,  \frac{\delta f_{W}^{v}}{\delta q_{\tau}^{h}} =  0 , \frac{\delta f_{W}^{v}}{\delta q^{v}} =-\frac{N\hat{W}^{h}(1-p^{v})}{\tau_{2}^{2}C}-\frac{\lambda (\hat{W}^{h})^{2}}{2N\tau_{2}^{2}C},  \frac{\delta f_{W}^{v}}{\delta q_{\tau}^{v}} = 0.$}
		\end{align*}
		
	\end{description}
	
	The linearised queue dynamic can be written in the form,
	\begin{eqnarray}\label{A.5}
		\frac{\partial \delta q^{h}(t_{1},t_{2})}{\partial t_{1}}&=& \frac{\delta f_{q}^{h}}{\delta W^{h}} \delta W^{h}(t_{1},t_{2}) + \frac{\delta f_{q}^{h}}{\delta W_{\tau}^{h}} \delta W^{h}(t_{1}-\tau(t_{1}),t_{2}) \nonumber\\
		&&\hspace{-1pt}+\hspace{-1pt} \frac{\delta f_{q}^{h}}{\delta W^{v}} \delta W^{v}(t_{1},t_{2}) \hspace{-1pt}+\hspace{-1pt} \frac{\delta f_{q}^{h}}{\delta W_{\tau}^{v}} \delta W^{v}(t_{1},t_{2}-\tau(t_{2})) \nonumber\\
		&&\hspace{-1pt}+\hspace{-1pt} \frac{\delta f_{q}^{h}}{\delta p} \delta p^{h}(t_{1},t_{2}) 
		\hspace{-1pt}+\hspace{-1pt} 	\frac{\delta f_{q}^{h}}{\delta p_{\tau}} \delta p^{h}(t_{1}-\tau(t_{1}),t_{2}) \nonumber\\
		&& \hspace{-1pt}+\hspace{-1pt}\frac{\delta f_{q}^{h}}{\delta p} \delta p^{v}(t_{1},t_{2}) 
		\hspace{-1pt}+\hspace{-1pt} 	\frac{\delta f_{q}^{h}}{\delta p_{\tau}} \delta p^{v}(t_{1},t_{2}-\tau(t_{2}))\nonumber\\
		&& \hspace{-1pt}+\hspace{-1pt} \frac{\delta f_{q}^{h}}{\delta q^{h}} \delta q^{h}(t_{1},t_{2})  
		\hspace{-1pt}+\hspace{-1pt} \frac{\delta f_{q}^{h}}{\delta q_{\tau}^{h}} \delta q^{h}(t_{1}-\tau(t_{1}),t_{2})\nonumber\\
		&& \hspace{-10pt}+\hspace{-1pt} \frac{\delta f_{q}^{h}}{\delta q^{v}} \delta q^{v}(t_{1},t_{2}) 
		 \hspace{-1pt}+\hspace{-1pt} \frac{\delta f_{q}^{h}}{\delta q_{\tau}^{v}} \delta q^{v}(t_{1},t_{2}-\tau(t_{2})) \hspace{-1pt}+\hspace{-1pt} h.o.t, \nonumber \\
		\frac{\partial \delta q^{v}(t_{1},t_{2})}{\partial t_{2}}&=& \frac{\delta f_{q}^{v}}{\delta W^{h}} \delta W^{h}(t_{1},t_{2}) \hspace{-1pt}+\hspace{-1pt} \frac{\delta f_{q}^{v}}{\delta W_{\tau}^{h}} \delta W^{h}(t_{1}-\tau(t_{1}),t_{2}) \nonumber\\
		&& \hspace{-1pt}+\hspace{-1pt}\frac{\delta f_{q}^{v}}{\delta W^{v}} \delta W^{v}(t_{1},t_{2}) \hspace{-1pt}+\hspace{-1pt} \frac{\delta f_{q}^{v}}{\delta W_{\tau}^{v}} \delta W^{v}(t_{1},t_{2}-\tau(t_{2})) \nonumber\\
		&&\hspace{-1pt}+\hspace{-1pt} \frac{\delta f_{q}^{v}}{\delta p} \delta p^{h}(t_{1},t_{2}) \hspace{-1pt}+\hspace{-1pt} 	\frac{\delta f_{q}^{v}}{\delta p_{\tau}} \delta p^{h}(t_{1}-\tau(t_{1}),t_{2})  \nonumber\\
		&& \hspace{-1pt}+\hspace{-1pt}\frac{\delta f_{q}^{v}}{\delta p} \delta p^{v}(t_{1},t_{2}) \hspace{-1pt}+\hspace{-1pt} 	\frac{\delta f_{q}^{v}}{\delta p_{\tau}} \delta p^{v}(t_{1},t_{2}-\tau(t_{2})) \nonumber \\
		&& \hspace{-1pt}+\hspace{-1pt} \frac{\delta f_{q}^{v}}{\delta q^{h}} \delta q^{h}(t_{1},t_{2})  \hspace{-1pt}+\hspace{-1pt} \frac{\delta f_{q}^{v}}{\delta q_{\tau}^{h}} \delta q^{h}(t_{1}-\tau(t_{1}),t_{2}) \nonumber\\
		&& \hspace{-20pt}+\hspace{-1pt} \frac{\delta f_{q}^{v}}{\delta q^{v}} \delta q^{v}(t_{1},t_{2})  \hspace{-1pt}+\hspace{-1pt} \frac{\delta f_{q}^{v}}{\delta q_{\tau}^{v}} \delta q^{v}(t_{1},t_{2}-\tau(t_{2}))\hspace{-1pt}+\hspace{-1pt} h.o.t,
	\end{eqnarray}
	where  $ f_{q}^{h} := 	\frac{\partial q^{h}(t_{1},t_{2})}{\partial t_{1}} $ and $ f_{q}^{v} := 	\frac{\partial q^{v}(t_{1},t_{2})}{\partial t_{2}} $. The partial derivatives that describes the queue dynamic are as,
	\begin{description}
		\item[ECN OFF:] Horizontal dimension
		\begin{align*}
			&\frac{\delta f_{q}^{h}}{\delta W^{h}} = \frac{N(1-\hat{p}^{v})}{\hat{\tau}_{1}},
			\frac{\delta f_{q}^{h}}{\delta W_{\tau}^{h}} =0,
			\frac{\delta f_{q}^{h}}{\delta W^{v}} = 0,\frac{\delta f_{q}^{h}}{\delta q_{\tau}^{h}} = 0 , \\
			&
			\frac{\delta f_{q}^{h}}{\delta W_{\tau}^{v}} = 0, \frac{\delta f_{q}^{h}}{\delta p^{h}} = 0 , \frac{\delta f_{q}^{h}}{\delta p_{\tau}^{h}} = 0,  \frac{\delta f_{q}^{h}}{\delta p^{v}} = 0, \frac{\delta f_{q}^{h}}{\delta q^{v}} =0,\\
			& \frac{\delta f_{q}^{h}}{\delta p_{\tau}^{v}} = -\frac{N\hat{W}^{h}}{\hat{\tau}_{1}}, \frac{\delta f_{q}^{h}}{\delta q^{h}} =\frac{N\hat{W}^{h}}{\tau_{1}^{2}C}(1-\hat{p}^{v}), \frac{\delta f_{q}^{h}}{\delta q_{\tau}^{v}} = 0,
		\end{align*}
		\item[ECN OFF:] Vertical dimension:
		\begin{align*}
			\frac{\delta f_{q}^{v}}{\delta W^{h}} &=0,
			\frac{\delta f_{q}^{v}}{\delta W_{\tau}^{h}} =0,
			\frac{\delta f_{q}^{v}}{\delta W^{v}} = \frac{N(1-p^{v})}{\hat{\tau}_{2}},
			\frac{\delta f_{q}^{v}}{\delta W_{\tau}^{v}} = 0, \\
			\frac{\delta f_{q}^{v}}{\delta p^{h}} &= 0 , \frac{\delta f_{q}^{v}}{\delta p_{\tau}^{h}} = 0,  \frac{\delta f_{q}^{v}}{\delta p^{v}} = -\frac{N\hat{W}^{v}}{\hat{\tau}_{2}} , \frac{\delta f_{q}^{v}}{\delta p_{\tau}^{v}} =0,\\
			\frac{\delta f_{q}^{v}}{\delta q^{h}} &=0,  \frac{\delta f_{q}^{v}}{\delta q_{\tau}^{h}} = 0 , \frac{\delta f_{q}^{v}}{\delta q^{v}} =\frac{N\hat{W}^{v}}{\tau_{2}^{2}C}(1-\hat{p}^{v}),  \frac{\delta f_{q}^{v}}{\delta q_{\tau}^{v}} = 0.
		\end{align*}
	\end{description}
\end{document}